\documentclass[12pt]{article}
\usepackage{hyperref} 
\usepackage{graphicx}
\usepackage{caption}
\usepackage{subcaption}
\usepackage{float}
\usepackage{amsmath}
\usepackage{style}
\usepackage{authblk}

\title{Towards Analyzing Formic  Acid Using Classical and Quantum Methods}
\date{March 23, 2026}

\author[1]{Omer Gurevich}
\author[1,2]{Tal Mor}
\author[1]{Ido Ram}

\affil[1]{Computer Science Department, Technion, Haifa, Israel\footnote{Email:  \texttt{omergu@campus.technion.ac.il}}}
\affil[2]{The Helen Diller Quantum Center, Technion, Haifa, Israel}

\begin{document}
\maketitle
\begin{abstract}
    Catalytic carbon fixation to formic acid is important for studying the
reduction of carbon footprint and the emergence of life. Can discrete quantum exhaustive search merged with other methods help reduce the carbon footprint?
We suggest merging quantum, quantum inspired, and classical tools for
a better simulation of various relevant processes. Quantum tools are
often used for analyzing the electronic structure of molecules, sometimes because
this problem is not scalable (in the number of orbitals) on classical
computers while it is potentially approximately scalable on (future)
quantum computers. It is potentially even solvable in the near future
using variational quantum eigensolvers (VQE) yet a major obstacle to
such analysis is the appearance of barren plateaus in the Hilbert
space describing the problem.

Here we make use of the basic (standard) tools while also including a
novel one --- the discrete quantum exhaustive search, which relies on
mutually unbiased bases, for analyzing the simplest non-catalytic
process involving carbon dioxide, hydrogen and formic acid.
\end{abstract}
\section{Introduction}
\label{sec:intro}

Quantum computers are expected to revolutionize fields like quantum chemistry by efficiently simulating many-body wavefunctions, potentially accelerating global decarbonization through the discovery of superior catalysts \cite{ref:von2021quantum}. In this work, we focus on Formic Acid, its implications on climate change, carbon fixation, and its connection to the origin of life. In what follows, we provide motivation for the significance of this interaction from both climatic and biological perspectives, and then describe the tools we used for studying it—tools from the worlds of both classical and quantum algorithms.

\subsection{Carbon fixation and climate change}

Anthropogenic carbon dioxide emissions are a major driver of global climate change~\cite{ref:ipcc2021,ref:nas2014}, and the development of sustainable technologies to capture and convert CO$_2$ into value-added products is a critical social and scientific challenge. One of the simplest and most studied reactions in this context is the hydrogenation of CO$_2$ to formic acid (HCOOH) \cite{ref:Moret2014}:
\[
\text{CO}_2 + \text{H}_2 \rightarrow \text{HCOOH}
\]
This reaction is deeply related to carbon fixation as it allows us to artificially transform inorganic CO$_2$ into organic and harmless formic acid.

Formic acid is already a valuable commodity chemical with a global production of about 800,000 tons annually, primarily used in textiles, cleaning, and preservation \cite{ref:Moret2014}. Its high volumetric hydrogen density ($53$ g/L of H$_2$), low toxicity, and liquid state under ambient conditions make it an attractive hydrogen storage medium \cite{ref:singh2016}. While hydrogen release from formic acid is well established, mainly using noble-metal catalysts, developing efficient base-free catalysts for the reverse reaction—hydrogenating CO$_2$ to formic acid—could significantly enhance its role in a future hydrogen economy \cite{ref:Moret2014} and establish it as a promising hydrogen carrier for energy storage and transport~\cite{ref:singh2016}. Efficiently modeling this process at the atomic level is important not only for fundamental understanding but also for guiding the rational design of catalysts that enable CO$_2$ utilization at scale.

\paragraph{Impact on humanity}
The uncatalysed CO$_2$ + H$_2$ $\rightarrow$ HCOOH reaction is slow and unfavorable when it is attempted in standard conditions. For this reason, the uncatalysed version is not considered useful for carbon fixation. However, a promising modern approach is to find catalysts which make the reaction more favorable by lowering its activation energy.
If humanity were to discover increasingly better catalysts, it could potentially increase the feasibility of reducing carbon emission by turning it to formic acid, and maybe even reverse some of the damage already caused to the Earth's climate and atmosphere.

\subsection{Formic acid and Origin of Life}
\begin{figure}[H]
    \centering
    \includegraphics[width=0.5\linewidth]{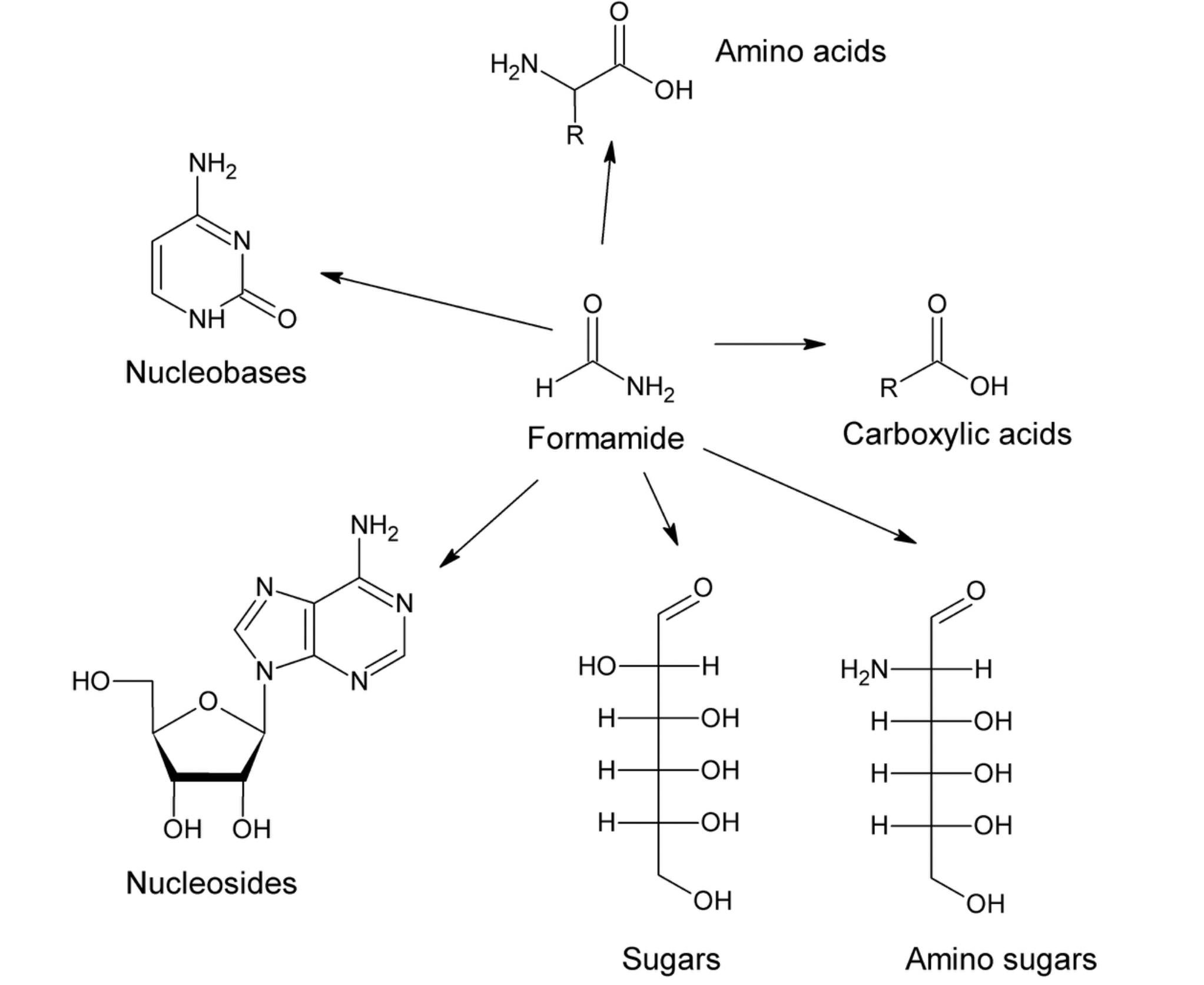}
    \caption{Main prebiotic building blocks that can be synthesized from formamide under plausible prebiotic conditions \cite{ref:saladino2012genetics}, \cite{ref:saladino2015meteorite}.}
    \label{fig:Formamide}
\end{figure}

While the role of formic acid in the origin of life is still not fully understood, it is often considered a possible precursor in prebiotic chemistry \cite{ref:demassa2013formic}.
Demassa et al.\cite{ref:demassa2013formic} further suggest, that there may have been an abundance of formic acid in the early Earth atmosphere. The exact effect of formic acid on abiogenesis is debated. In addition, they argue that formic acid could inhibit amino acid synthesis, while Mohammadi et al.\cite{ref:mohammadi2020formic} suggest that it could contribute indirectly by promoting the formation of formamide (HCONH$_2$)\cite{ref:mohammadi2020formic}. Formamide is widely regarded as a versatile prebiotic building block, capable of giving rise to amino acids, nucleobases, and sugars under appropriate conditions. As can be seen in \Fig{fig:Formamide} (taken from Wikipedia and based on \cite{ref:saladino2012genetics} and \cite{ref:saladino2015meteorite}), there are several prebiotic building blocks that can be synthesized from formamide under plausible prebiotic conditions, including amino acids.
Thus, formic acid may have acted either as a direct precursor or as a feedstock for other intermediates like formamide, which in turn drive Miller–Urey type chemistry. More generally, it might be that the process CO$_2$ + H$_2$ $\rightarrow$ HCOOH has established an equilibrium between CO$_2$, H$_2$, and formic acid under early Earth conditions, thereby seeding more complex prebiotic chemistry.

\subsection{Review of the Algorithms Used}
\label{sec:alg_rev}
In our study, we apply a combination of classical and quantum computational methods to analyze this reaction. For the quantum simulations, we investigate the Variational Quantum Eigensolver (VQE). Further details on VQE are provided in \Sec{sec:vqe}, where the whole hybrid scheme for molecular structure problems is detailed in \Sec{sec:hybrid}. Although VQE is a leading approach for molecular ground-state estimation, it often encounters the Barren Plateau (BP) problem \cite{ref:mcclean2018}. The BP problem occurs when cost-function gradients vanish exponentially with system size, rendering standard optimization ineffective \cite{ref:mcclean2018, ref:Larocca2025}. To mitigate this, we follow a recent scheme proposed by Alfassi et al. (2024) called \textit{Discrete Quantum Exhaustive Search} (DQES) \cite{ref:AMM2024}. DQES relies on initializing the optimization from states that \textit{evenly spread} across the Hilbert space, specifically utilizing states from Mutually Unbiased Bases (MUBs)\cite{ref:durt2010mutually} to avoid flat landscapes and local minima. Our results show that, potentially, DQES-based VQE is more precise than random initialization on average, indicating that barren plateaus and local minima may be successfully avoided in problem instances where standard VQE fails to converge.
\\
For the construction of potential energy surfaces (PES), we utilize quantum-inspired methods based on matrix diagonalization of qubit Hamiltonians \cite{ref:pople1999}. To characterize the full reaction coordinate and locate transition states, we employ classical optimization algorithms such as Nudged Elastic Band (NEB) \cite{ref:neb} and Intrinsic Reaction Coordinate (IRC) \cite{ref:irc}.

\section{Background}
In this section, we review the foundational concepts required to study chemical reactions through both classical and quantum lenses. Then we review the Barren Plateau problem in variational quantum algorithms. While we hope these quantum algorithms will be able to solve chemistry problems larger than those solvable classically, the existence of Barren Plateaus is currently viewed as a major obstacle in their scalability.  

\subsection{Electronic Structure Problem}
\label{sec:elec_struct}
One of the most basic tasks in computational chemistry is to determine the ground-state energy of a molecule or a system of several molecules. This is known as the \textit{electronic structure problem}. The input is the spatial coordinates of the atoms involved, sometimes along with other properties such as total charge and spin of the system. It is also possible to find excited energy levels of the system of interest. 

In an isolated atom, electrons occupy atomic orbitals (AOs)—continuous spatial functions describing the probability distribution of the electron's position. Each spatial AO corresponds to two spin-orbitals. While an atom theoretically possesses an infinite number of energy levels and corresponding AOs, thermodynamic considerations at finite temperatures dictate that only the lowest energy states are meaningfully occupied. Consequently, the physical system is well-approximated by considering only a finite number of AOs, typically not much more than half the number of electrons. When atoms interact to form a molecule, their individual AOs overlap to create chemical bonds. The electrons in the resulting molecular system are then described by molecular orbitals (MOs), which are mathematically constructed as linear combinations of the constituent atomic orbitals (LCAO). The multi-particle fermionic states, which must satisfy antisymmetry, are mathematically represented by Slater determinants within the Fock space. In \App{sec:chem_prelim} we provide a more formal background on these topics. 

Full Configuration Interaction (FCI) represents the gold standard for solving the electronic Schrödinger equation exactly within a given finite space. However, the dimension of the Fock space $d$ scales combinatorially as $\binom{2N}{N_e}$ for $N_e$ electrons in $N$ molecular orbitals, i.e $2N$ spin-orbitals. Such exponential scaling imposes hard limits on classical exact diagonalization, rendering it intractable beyond approximately $N_e=16$ electrons in $2N=32$ spin-orbitals \cite{ref:tennyson2022,ref:levine2020}. 

Despite this exponential growth, it is possible to simulate intermediate-size molecules by restricting the calculation to a limited subset of critically important MOs (e.g., those directly involved in bond breaking and formation) and electrons while freezing the rest. This is referred to as choosing an \textit{Active Space (AS)}. More details on how FCI is commonly used are explained in \Sec{sec:single_point}.

To avoid the exponential cost of FCI, classical quantum chemistry relies on various additional approximation methods. Among them, it relies on mean-field and density-based approximations. Hartree–Fock (HF) theory approximates ground-state energies via averaged electron-electron potentials, while Density Functional Theory (DFT) approximates electron correlation via exchange–correlation functionals. In the general worst-case scenario, solving the HF and DFT problems is still highly complex; as discussed later in \Sec{sec:complexity}, the HF problem is NP-complete, and the DFT problem is QMA-complete \cite{ref:schuch2009}.

\subsection{Potential Energy Surface}
\label{sec:PES}
When studying dynamics and reactions, scientists are interested in knowing the energy of the system at many points, that is, solving the electronic structure problem for many spatial configurations of the atoms. The plot of the energy as a function of the coordinates is called Potential Energy Surface (PES). The construction of the PES lies at the core of modeling chemical reactions. The PES can give for example insights about stable and non-stable regions of spatial configurations, as well as the likely reaction trajectory at any given point.

We present two examples at \Sec{sec:PESRes} on collinear systems - one of three hydrogen atoms, and the second of CO$_2$. These are simple examples that pave the way in studying our CO$_2$ + H$_2$ $\rightarrow$ HCOOH reaction of interest. 

\subsection{Reaction Coordinate}
\label{sec:RC}
The chemical reaction process starts with reactants and ends with products. The abstract set of coordinates along the reaction path is called the reaction coordinate. Mathematically, the reaction coordinate is a path of points that have minimum energy in all directions perpendicular to the path connecting the reactants and the products. In simple systems with only one or two degrees of freedom, plotting the PES can be the first step for finding the Reaction Coordinate, whereas for reactions with more complex geometries, other methods should be considered. We discuss optimization techniques that are used in such cases in \Sec{sec:IRC}.\\

\subsection{Transition States}
Once the reaction coordinate is known, there is a specific point of interest along this path called Transition State. Out of all the points on the reaction coordinate, it is defined as the point of maximal energy - the 'bottleneck' of the reaction. On the PES, it is a saddle point: It has maximal energy along the reaction coordinate, and minimal energy perpendicular to it. Discovery of transition states of reactions is very useful as it teaches us the activation energy required for the reaction to happen, and guides us in the search for better catalysts. 

\section{Classical Methods}
\subsection{Single-Point Energy Calculation}
\label{sec:single_point}
All the computational methods mentioned in \Sec{sec:elec_struct} rely on representing AOs by using a finite basis set of functions. Hence, the $N$ MOs are essentially a linear combination of these $N$ basis functions rather than of exact AOs. Typically, $N$ scales linearly with the number of electrons $N_e$~ \cite{ref:jensen2017introduction}. \\
Because modeling exact AOs requires deep chemical and physical considerations, it is standard practice to rely on predefined basis sets developed by domain experts. 
Two well-known basis sets are STO-3G and 6-31G. In the STO-3G basis set, a basis function is assigned to every AO up to the highest energy level occupied by the electrons  \cite{ref:jensen2017introduction}. Each basis function is built from $3$ Gaussian functions called primitives (where more generally basis sets STO-nG use $n$ Gaussians per basis function) , resulting in an approximation for the AO which is computationally efficient to integrate. More expressive basis sets assign multiple functions per AO to increase accuracy. For example, the 6-31G basis set uses two basis functions for each valence orbital, correspondingly increasing $N$ and the overall computational complexity, where the basis functions are also approximated by a combination of Gaussian functions.  Consider for example the atoms relevant for carbon fixation:  In STO-3G (6-31), each hydrogen contributes 1 (2), and each carbon/oxygen contributes 5 (9) basis functions. Throughout this work, we exclusively utilize the STO-3G.
By limiting the problem to a small number of basis functions $N$ (which scales linearly in the input size), standard implementations of HF and DFT can utilize iterative procedures that exhibit favorable polynomial scaling per iteration (e.g., $\mathcal{O}(N^3)$--$\mathcal{O}(N^7)$ \cite{ref:jensen2017introduction}). Note that exact methods such as FCI are only exact within a given basis set, and if the active space is full, meaning it includes all $N$ molecular orbitals. Restricting the active space—typically by freezing lower-energy core electrons, as mentioned in \Sec{sec:elec_struct}—further reduces the accuracy of the calculation.

Although this polynomial efficiency per iteration is nice, it is insufficient in the presence of strong electronic correlation---such as near avoided crossings, conical intersections, or in catalytic complexes involving transition metals or charge transfer states. In these hard cases, the underlying exponential worst-case complexity of HF and DFT manifests directly: the number of iterations can grow exponentially in $N$, or convergence may not be reached at all \cite{ref:lyakh2012CC,ref:luthi1982HF,ref:cai2002failureDFT}.

For catalytic CO\(_2\) fixation, the complexity arises not merely from system size but from the need to capture subtle multireference character, electron delocalization, and dynamically evolving correlations during bond formation and breaking. Such effects challenge single-reference methods and preclude classical black-box modeling of reaction pathways involving heterogeneous catalysts.

In this work, we use HF as an initial approximation. From this reference state, constructing and subsequently solving the second-quantized Hamiltonian constitutes what is called the post-Hartree–Fock (post-HF) process. Finding its eigenvalue is done either quantumly (\Sec{sec:DQES}) or in a quantum-inspired manner (\Sec{sec:PESRes}); both equivalent to solving FCI. We elaborate on the process starting from a classical approximation to quantumly finding the exact eigenvalue in \Sec{sec:hybrid}.

\subsection{Potential Energy Surface Calculation}
By using the above-mentioned methods for solving the electronic structure problem, a PES can be obtained simply by finding the energy for points on a multidimensional grid representing the coordinates of the atom. For systems with more than 2 degrees of freedom, a visual representation becomes challenging, but it is still possible to study the PES. 

For the uncatalyzed CO$_2$ + H$_2$ $\rightarrow$ HCOOH reaction, the PES is relatively simple. Classical methods can locate the transition state and compute the activation barrier and reaction pathway with reasonable accuracy. However, when a \textit{catalyst} is introduced, the complexity of the system increases dramatically: the number of atoms grows, the electronic structure becomes more strongly correlated, and the PES develops a rugged, high-dimensional landscape with many local minima and competing pathways. Capturing the role of the catalyst — and the transition states and intermediates it enables — often requires going beyond mean-field approximations to include dynamic correlation and multireference character, which is where classical methods struggle.

For this work, we do not introduce a catalyst. Still, we consider alternatives to calculating the entire PES. We discuss these alternative ways to study reaction coordinates and transition states in the upcoming sections \ref{sec:NEB} and \ref{sec:IRC}.

\subsection{Transition State Calculation - NEB Method}
\label{sec:NEB}
Due to the importance of finding a valid transition state, there are many algorithms with varying advantages and disadvantages for finding the transition states of various reactions. In this paper we used the Nudged Elastic Band (NEB) method \cite{ref:neb}. The method approximates the reaction coordinate by assigning $N$ points on the PES that connect between the reactants and the products. Then, it iteratively optimizes each point according to the PES (that is, to minimize its energy) but also accounts for 'elastic forces' between consecutive points which ensures the path taken won't collapse to a trivial point. This way, the NEB algorithm finds a path that is minimal in all axes (except tangentially to the path) and connects between the reactants and products. The highest-energy point on this path is thus an approximation of the transition state, which can be taken as-is or further optimized by iterative optimization methods that rely on good guesses of the transition state.    

\subsection{Intrinsic Reaction Coordinate}
\label{sec:IRC}
To calculate the reaction coordinate more precisely, we use the IRC method \cite{ref:ircalgo, ref:irc}. The algorithm begins from the transition state, nudges it in the direction downstream in the energy (this is done by considering the lowest frequency eigenvector of the hessian, known as the "transition vector", at the transition state), and then iteratively optimizes the geometry until it reaches the minimum, which should be the reactants or the products. Then it re-optimizes in the opposite direction to the 'transition vector' until it reaches the reactants or the products. Finally, stitching the obtained backward and forward paths together yields the full reaction coordinate of the reaction.  In cases where there are intermediate products, or multiple local transition states, one needs to run the method multiple times from each local transition state and stitch the full path together.
\section{Quantum Methods}
\subsection{Variational Quantum Eigensolver (VQE)}
\label{sec:vqe}
The Variational Quantum Eigensolver (VQE)~\cite{ref:peruzzo2014,ref:tilly2022a} is a prominent example of the broader class of Variational Quantum Algorithms (VQAs). VQAs are iterative hybrid algorithms where a classical optimizer trains a Parametrized Quantum Circuit (PQC), denoted as $U(\boldsymbol{\theta})$, with $\boldsymbol{\theta}$ representing a vector of tunable parameters. In VQE specifically, the ansatz $U(\boldsymbol{\theta})$ prepares a trial quantum state $\ket{\Psi(\boldsymbol{\theta})}$ in the Hilbert space of $n$ qubits, with the objective of approximating the ground state energy of a physical system's Hamiltonian $\hat H$. For the sake of concreteness, \App{sec:Hams} provides the explicit Hamiltonians describing some of the molecules and chemical reactions simulated in our experiments.

The cost function is defined as the expectation value of this Hamiltonian:
\begin{equation}
\label{eq:cost}
    C(\boldsymbol{\theta}) = \obraket{\Psi(\boldsymbol{\theta})}{\hat H}{\Psi(\boldsymbol{\theta})},
\end{equation}
where $\ket{\Psi(\boldsymbol{\theta})} \coloneqq U(\boldsymbol{\theta})\ket{\Psi_0}$. Often, the initial state is set to the computational basis state $\ket{\Psi_0}=\ket{0}^{\otimes n}$. However, this is not always the case; for example, in problems from quantum chemistry, the process frequently starts from a physically inferred state, such as the Hartree-Fock (HF) state (see Sec.~\ref{sec:chem_bg}).

To calculate the cost, $\hat H$ is decomposed into a linear combination of Pauli strings, $\hat H = \sum_P c_P P$, where $c_P$ are real coefficients and $P \in \{I, X, Y, Z\}^{\otimes n}$\footnote{For reference, the standard Pauli matrices are defined as: 
$I=\left(\begin{smallmatrix}1&0\\0&1\end{smallmatrix}\right)$, 
$X=\left(\begin{smallmatrix}0&1\\1&0\end{smallmatrix}\right)$, 
$Y=\left(\begin{smallmatrix}0&-i\\i&0\end{smallmatrix}\right)$, 
$Z=\left(\begin{smallmatrix}1&0\\0&-1\end{smallmatrix}\right)$. 
The Hadamard gate and phase gate that we mention below are $H=\frac{1}{\sqrt{2}}\left(\begin{smallmatrix}1&1\\1&-1\end{smallmatrix}\right)$ and $S=\left(\begin{smallmatrix}1&0\\0&i\end{smallmatrix}\right)$, respectively.}. For quantum chemistry Hamiltonians of $N$ molecular orbitals, the number of Pauli strings is $O(N^4)$ \cite{ref:tilly2022a}. By linearity:
\begin{equation}
    C(\boldsymbol{\theta}) = \sum_P c_P \obraket{\Psi(\boldsymbol{\theta})}{P}{\Psi(\boldsymbol{\theta})}
\end{equation}
Since measurement is performed in the computational basis, measuring $X$ or $Y$ operators on a given qubit requires applying basis rotation gates to map their eigenstates to the computational basis. For a Pauli-$X$ operator, a Hadamard gate $H$ is applied. For a Pauli-$Y$ operator, a phase gate $S^\dagger$ followed by $H$ is applied. These measurements are repeated to obtain statistically reliable estimates, which are summed to compute $C(\boldsymbol{\theta})$. A classical optimizer then determines the parameter update for the next iteration. \\
As you can see, at each iteration, every Pauli term has to be measured repeatedly enough times to get a reasonable approximation of its expectation value. As was mentioned in the Introduction (\Sec{sec:alg_rev}), our full quantum method is to apply DQES-based VQE, in which the optimization starts from different MUB states. When performing DQES-based VQE, all the measurements have to be repeated for each of the initial MUB states.

The definition in Eq.~\eqref{eq:cost} can be extended to mixed states by defining $C(\boldsymbol{\theta})=\Tr\left(\rho(\boldsymbol{\theta})\hat H\right)$, though we restrict our discussion here to pure states. Finally, as noted in the Introduction, the optimization of these landscapes is non-trivial; VQAs frequently suffer from Barren Plateaus (BPs), where gradients vanish exponentially with the system size, hindering convergence. We now turn to discuss these BPs in more detail.
\subsection{Barren Plateaus}
\label{sec:BP}
For the optimization using VQEs, Barren plateaus (BPs) usually constitute a major obstacle. More specifically, an ansatz is considered suitable for optimization when its cost-function gradient decreases at most polynomially with system size or circuit depth; conversely, when the gradient decays exponentially, the algorithm is said to suffer from a barren plateau \cite{ref:tilly2022a,ref:Larocca2025}. From a statistical viewpoint, this phenomenon emerges because the expectation value of gradients necessary for the iterations for the optimization problem concentrates exponentially around zero \cite{ref:mcclean2018}, and as the number of qubits grows, the variance also vanishes. In other words, the probability of selecting a random state far from a region of gradient zero decreases severely with the size of the system. In this sense, even highly expressive ansätze which explore the Hilbert space almost uniformly are extremely unlikely to sample parameters with non-negligible gradients. This effect is not limited to gradient-based methods: it has been shown to also hinder gradient-free optimizers such as COBYLA, Powell, and Nelder-Mead \cite{ref:tilly2022a}.

Several mitigation strategies have been proposed to address BPs. Techniques that reduce ansatz expressibility such as adaptive ansätze help avoid exponentially vanishing gradients \cite{ref:tilly2022a}. In this direction, initializing parameters in regions of high gradient--analogous to positioning near ``steeper gravitational wells''--could improve optimization by avoiding flat regions. However, as noted in Ref. \cite{ref:mcclean2018}, due to measure concentration, such favorable regions occupy exponentially small volumes of the parameter space and are difficult to detect.
\\
Another leading strategy was proposed by Grant et al. \cite{ref:Grant2019}. This strategy is more similar to ours as both strategies deal with alternative initialization. Grant et al. suggest to choose initial parameters such that the circuit approximates the identity operation or a block-identity structure, preventing the gradients from vanishing when training starts. However, problem may still remain; the parameters may be initialized near local traps (despite having large gradients) or within a classically simulable subspace~\cite{ref:cerezo2022}. 
\section{Hybrid Quantum-Classical Computational Pipeline in VQE}
\label{sec:hybrid}
\subsection{From Geometry to Qubits: The Workflow}
\label{sec:chem_bg}
Our research utilizes a hybrid quantum-classical workflow mediated by Qiskit Nature~\cite{ref:qiskit_nature} and the classical chemistry driver PySCF~\cite{ref:pyscf}. The transformation from a physical molecule to a qubit Hamiltonian involves four computational stages:

\begin{enumerate}
    \item \textbf{System Definition:} The input is defined by atomic coordinates, total charge, and spin multiplicity. We assume the standard Born-Oppenheimer approximation (see Eq. \ref{eq:elec_ham} in \App{sec:chem_prelim}), treating nuclei as fixed point charges.
    
    \item \textbf{Basis Set \& Mean-Field Calculation:} We select a Basis Set (e.g., STO-3G or 6-31G) to represent the Atomic Orbitals (AOs), denoted as $\{\chi_\mu\}$. Let $N$ be the number of spatial basis functions.
    \begin{itemize}
        \item \textit{Integral Generation:} First, the classical driver computes the one-body and two-body integrals in the AO basis $\{\chi_\mu\}$. There are $O(N^4)$ such integrals, representing all possible Coulombic interactions between the basis functions. The form of these integrals is given in Eqs. \ref{eq:1body},
        \ref{eq:2body} in \App{sec:chem_prelim}.
        \item \textit{Hartree-Fock (HF) Optimization:} We perform a Self-Consistent Field (SCF) calculation~\cite{ref:szabo_ostlund} to find the Molecular Orbitals (MOs), denoted as $\{\psi_p\}$. The MOs are linear combinations of AOs:
\begin{equation}
    \psi_p = \sum_{\mu=1}^N C_{\mu p} \chi_\mu,
\end{equation}
where $\{x_\mu\}$ are the basis functions chosen to represent the AOs. 
The SCF algorithm iteratively optimizes the coefficient matrix $C$ by solving the \textit{Hartree-Fock equations}. This requires constructing the \textit{Hartree-Fock potential}~\cite{ref:szabo_ostlund} at each iteration (at a cost of $O(N^4)$) to match the updated value of $C$, until convergence.
Full mathematical details of the HF equations and potential are provided in \App{sec:hf_method}. Formally, this heuristic procedure attempts to solve the \textit{Hartree-Fock problem}~\cite{ref:schuch2009}; we discuss the formal definition and complexity of this problem immediately below.
\item \textit{Output of step 2:} Once the optimized coefficient matrix $C$ is found, the original AO integrals are transformed (rotated) into the MO basis via a tensor contraction ($C$ is the basis transformation matrix from AOs to MOs). This transformation is given by:
        \begin{equation}
        \label{eq:hpqrs}
             h_{pqrs} = \sum_{\mu \nu \lambda \sigma} C_{\mu p}^* C_{\nu q}^* C_{\lambda r} C_{\sigma s} \langle \mu \nu | \lambda \sigma \rangle_{AO},
        \end{equation}
        with
        \begin{equation}
    \langle \mu \nu | \lambda \sigma \rangle_{AO} = \int d\mathbf{x}_1 d\mathbf{x}_2 \, \chi_\mu^*(\mathbf{x}_1) \chi_\nu^*(\mathbf{x}_2) \frac{1}{|\mathbf{r}_1 - \mathbf{r}_2|} \chi_\lambda(\mathbf{x}_1) \chi_\sigma(\mathbf{x}_2)
\end{equation}
        \end{itemize}
        The resulting tensor $h_{pqrs}$ contains not only the Coulomb and Exchange terms used in HF but also the general off-diagonal terms required to describe electron correlation (excitations) in the VQE Hamiltonian which is obtained in the next step. Although diagonalizing the Hamiltonian takes exponential time, its construction is polynomial: A naive evaluation of all 4-index integrals (Eq.~\ref{eq:hpqrs}) scales as $O(N^8)$. Standard implementations perform this as a sequential tensor contraction (transforming one index at a time), reducing the cost to $O(N^5)$.
        \item \textbf{Active Space Transformation:} To make the problem tractable on near-term quantum hardware, we truncate the orbitals to an Active Space (AS), effectively freezing core electrons and ignoring high-energy virtual orbitals.
    \begin{itemize}
        \item \textit{Output of Step 3:} This step produces a reduced set of integrals ($h_{pq}^{AS}, h_{pqrs}^{AS}$) corresponding only to the active orbitals. These reduced integrals constitute the coefficients of the final Second Quantized Hamiltonian (Eq. \ref{eq:second_quant} in \App{sec:chem_prelim}). 
    \end{itemize}
    
    \item \textbf{Qubit Mapping:} The fermionic Hamiltonian obtained at the previous step, is projected onto the qubit space using transformations such as Jordan-Wigner~\cite{ref:jordan_wigner} or Parity~\cite{ref:Bravyi2002, ref:seeley_bravyi}. This produces a qubit operator $H_q = \sum_i c_i P_i$, composed of Pauli strings $P_i$, which serves as the input for the VQE algorithm. For example, see Hamiltonians (\ref{eq:H_H2O}) and (\ref{eq:H_formic}) in \App{sec:Hams}.
\end{enumerate}

\subsection{Benchmarking with Classical Solvers}
To validate the accuracy of our VQE result, we compare our results against the Full Configuration Interaction (FCI) energy~\cite{ref:sherrill_fci}. In our workflow, this benchmark is computed using Qiskit Nature's exact eigensolver, which performs matrix diagonalization of the Second Quantized Hamiltonian (defined in Eq. \ref{eq:second_quant} in \App{sec:chem_prelim}) within the chosen active space. 

This result is exact within the chosen basis set and within the choice of the active space. However, the calculation scales exponentially with system size, creating a strict computational bottleneck.
\\ 

\textit{Note on computational cost:} While Qiskit Nature's default solver relies on full diagonalization (scaling as $O(D^3)$ where $D$ is the full Hilbert space dimension), specialized chemistry drivers like PySCF offer iterative subspace methods such as the Davidson~\cite{ref:davidson1975} or Lanczos~\cite{ref:lanczos1950} algorithms. These methods reduce the cost to roughly $O(k D^2)$ (where $k$ is the number of desired eigenvalues), allowing for slightly larger systems, though they remain exponentially costly in the number of qubits.
\subsection{Remarks on Invariance and Complexity}
\label{sec:furtherRemarks}

    \subsubsection{Independence of FCI from Mean-Field Method:} PySCF offers various mean-field approximations, including Restricted HF, Unrestricted HF, and Density Functional Theory (DFT)~\cite{ref:parr_yang_dft}. While the details on each of these specific method is beyond our scope, a common point of confusion is whether the final FCI result depends on this initial choice.

The FCI wavefunction $|\Psi_{FCI}\rangle$ is defined as the linear combination of the reference state $|\Phi_0\rangle$ obtained from the chosen approximation method (e.g, the HF state obtained from the SCF procedure), and all possible excitations:
\begin{equation} \label{eq:fci_expansion}
    |\Psi_{FCI}\rangle = c_0 |\Phi_0\rangle + \sum_{ia} c_i^a |\Phi_i^a\rangle + \sum_{ijab} c_{ij}^{ab} |\Phi_{ij}^{ab}\rangle + \dots
\end{equation}
The term $|\Phi_i^a\rangle$ represents a singly excited determinant formed by exciting an electron from the occupied orbial $i$ to an unoccupied orbital $a$, $|\Phi_{ij}^{ab}\rangle$ represents a doubly excited determinant formed by exciting two electrons from occupied orbitals $i,j$ to unoccupied orbitals $a,b$, and so on... The orbitals mentioned are MOs.

The set of all such determinants forms a complete basis for the $N$-electron Hilbert space defined by the atomic basis set which we started with. Changing the initial method (e.g., from HF to DFT) essentially rotates the orbital basis; while the individual coefficients $c$ change to accommodate the new basis, the total vector space spanned by the determinants remains identical. Note that $\ket{\Phi_0}$ itself, and its excitations depend on the choice of approximation. But this is merely a change of basis. Thus, the exact diagonalization yields a unique ground state energy, independent of the initial orbitals.\footnote{Strictly speaking, this invariance holds only if the calculation is performed on the full basis set. If an Active Space is used, the results will agree only if the selected active orbitals span the exact same subspace.}
        \subsubsection{Feynman's Intuition and Beyond}
    \label{sec:complexity}
    Before connecting computational complexity to specific computational chemistry problems, we briefly clarify the relevant complexity classes. The class NP encompasses problems that a classical computer can solve in polynomial time, provided a correct and relevant hint is supplied by an external source. Analogously, the QMA (Quantum Merlin Arthur) complexity class contains problems that a quantum computer can solve in polynomial time once an appropriate hint is given from an outside source.
    While the HF algorithm seeks to find the optimal Slater determinant for minimizing the energy, a formulation of the promise (decision) HF problem for a system with $N$ spatial orbitals is as follows: given two constants $b>a$ with a gap $b-a > \frac{1}{\text{poly}(N)}$, does the optimal Slater determinant have an energy lower than $a$ or larger than $b$? This problem turns out to be NP-complete.
    Similarly, a decision problem for the DFT method is known to be QMA-complete. For both results, see~\cite{ref:schuch2009}.
\\
\newline
We now give the intuition for why the HF problem is NP-complete. The proof from \cite{ref:schuch2009} is based on the fact that the promise problem of whether an Ising spin glass Hamiltonian $H_{\text{ISING}}$ on an $L \times L \times 2$ lattice has ground state energy $e_g < a$ or $e_g > b$ is NP-complete~\cite{barahona1982computational}. It is possible to map any instance $(H_{\text{ISING}}, a, b)$ to an instance $(H_{\text{HF}}, a', b')$ in polynomial time such that:
\begin{itemize}
    \item If $e_g < a$, then $\min_{\Phi} \langle \Phi | H_{\text{HF}} | \Phi \rangle < a'$, where the minimization is over all Slater determinants $\Phi$.
    \item If $e_g > b$, then $\min_{\Phi} \langle \Phi | H_{\text{HF}} | \Phi \rangle > b'$.
\end{itemize}
Since the mapping can be done in polynomial time, the Hartree-Fock problem is NP-hard. Additionally, the constructed $H_{\text{HF}}$ is in second-quantized form (with only bi-quadratic terms; see Eq.~\ref{eq:second_quant} for an example of such Hamiltonian), which makes the evaluation for the energy of a witness state $\Phi$ efficient. Therefore, we conclude that the Hartree-Fock problem is NP-complete.
\\

Since the HF problem is NP-C, here is why we expect a quantum
advantage. Most problems in the class NP-C are not hard on average,
but in the worst cases. This leaves a lot of room of easy cases, but
not only. It also leaves a lot of room for cases that are potentially
easy for a quantum computer while they are still hard for classical
ones.  Intuitively, the quantum advantage relevant for those cases
could result from expanding the search space far beyond the
computation basis. Even for just a single qubit, the two values
accessible classically are generalized into a whole sphere. So one may
assume that some BP and local minima could potentially be bypassed by
alowing the huge space that expands beyond just the computation basis.
On small-size problems MUBs could provide non-conventional hints to
where such bypasses could be enabled. A similar intuition is not only
about easy cases but also about easy approximations. More about this,
here below.

When going beyond HF, this enlarged space is actually what is also accessible to Nature itself. In quantum terminology, instead of an Ising model, the correct and relevant model is the Heisenberg model. For context, finding the ground state of an Ising model is NP-complete~\cite{barahona1982computational}, as reviewed by Lucas~\cite{ref:Lucas_2014}, who formulated numerous classical combinatorial problems as the minimization of Ising Hamiltonian energies. The Heisenberg model, however, connects directly to the quantum Local Hamiltonian problem (where molecular Hamiltonians sit), which Kitaev established as being QMA-complete~\cite{ref:Kitaev2002Classical}. As a result, there is an even better intuition as to why quantum computers may be superior to classical ones: algorithms like VQE are motivated precisely by this fact, as they are designed to variationally prepare and explore these highly correlated states directly on quantum hardware, bypassing the classical memory bottlenecks that make such QMA-hard spaces intractable.

Feynman suggested the notion of quantum computers exactly for that
reason: Since Nature solves the problem of directing a molecule into
its ground state, and since Nature is quantum, maybe a quantum
computer can solve this problem exponentially faster than classical
ones?  However, today we know that the best estimate for solving the
ground state problem is using DFT instead of HF. But we also know that
DFT is even a harder problem than HF, and it is harder than being
NP-C. It is QMA-C.

So why can we expect Nature to solve QMA-C problems? Our intuition in this case is less about finding easy
cases that are easy for a quantum computer but hard for a classical
ones. Our intuition---going beyond Feynman's---says that Nature, for the hard cases, finds
great approximations instead of the optimal solution. So we are assuming
there are approximations to certain levels that are easy for a quantum
computer but are still hard for classical computers. This, in our
eyes, is the improved variant for what Feynman predicted.
\section{Results}
\label{sec:results}
\subsection{Quantum Inspired Analysis}
\label{sec:PESRes}
In \Sec{sec:PES} we reviewed the motivation for calculating the PES for reactions or molecules. We also discussed the complexity of doing so for large systems, namely large in the number of active orbitals. At this stage, we present calculations for smaller systems: The first is for the dynamics of 3 hydrogen atoms, which is a simple case study which demonstrates expected phenomena, which we discuss shortly. Two more systems are the CO$_2$ and H$_2$O molecules, both related to HCOOH. Carbon-dioxide (CO$_2$), is a relevant starting point towards our ultimate goal of studying PES related to carbon fixation, as it is involved in the process we study. We further investigate the dynamics of this process in following sections. 
Water is also relevant, as it is one of the reactants in another process where formic acid is formed (see \Sec{sec:discussion}).

The calculation is quantum inspired. We used the PySCF software that generates a fermionic Hamiltonian of the reaction, based on (restricted) Hartree-Fock approximation with the basis set STO-3G. Then, we used the parity mapping, to obtain a corresponding qubit Hamiltonian (e.g Eqs. (\ref{eq:H_H2O}) and (\ref{eq:H_formic})), which is done by  Qiskit-Nature (PySCF is also called from Qiskit-Nature). The energies calculated are the minimal eigenvalues of these Hamiltonians, calculated by matrix diagonalization, possible only for small number of qubits $n$, as the matrix size is exponential in $n$. \\

\begin{figure} [H]
\centering
\includegraphics[width=1.0\linewidth]{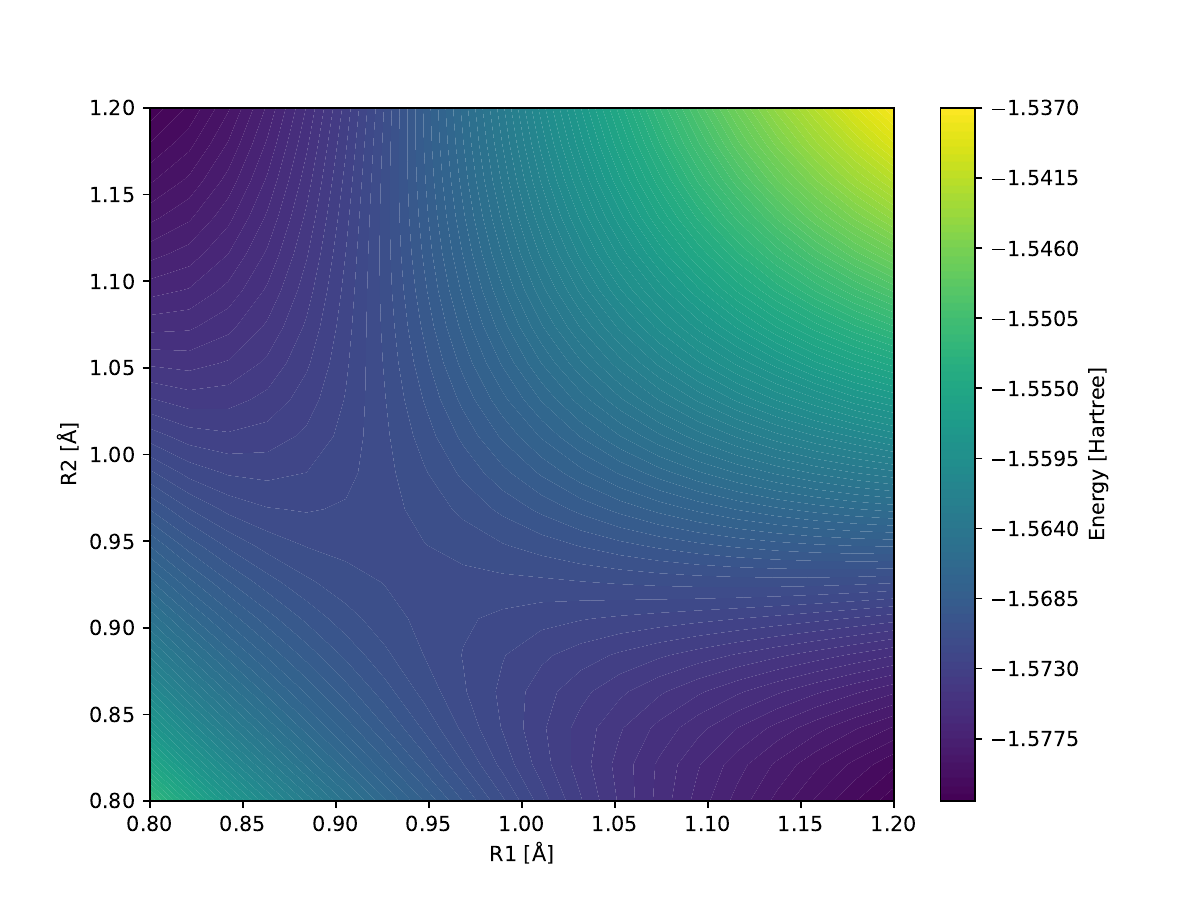}
    \caption{2d PES of co-linear dynamics of three hydrogen atoms.}
    \label{fig:H3}
\end{figure}

In the dynamical simulation of the three-hydrogen system, we assumed the three atoms are colinear and defined $R_1 [\text{\AA}]$ and $R_2[\text{\AA}]$  to be the distances between the atom in the center, to the left atom and to the right atom, respectively. For a 20 $\times$ 20 grid of evenly spread points $R_1,R_2 \in [0.8,1.2]$, we calculated the energy of the system. We chose an Active Space (AS) of three electrons and three spatial orbitals (six spin orbitals), which is the default in PySCF. The active space refers to a reduced subset of electrons and molecular orbitals selected for explicit treatment in qluantum chemical calculations, allowing accurate simulations at a fraction of the full Hilbert space size. All calculations were performed in the zero-temperature limit ($T = 0$~K), where thermal excitations are neglected and the system is characterized by its ground state. Accordingly, we compute the full spectrum of the qubit Hamiltonian and report the lowest eigenvalue as the ground state energy. The results for the potential energy surface (PES) of the system are shown in Figure~\ref{fig:H3}.


Our results exhibit the known behavior: Formation of H$_2$ molecule, in which two of the atoms are very close, and the third is not bound to them,  is the most preferable state. Indeed, we see two global minima of the energy in the upper left and bottom right regions of the figure. The formation of H$_3$ molecule as seen, is less preferable, though it appears as a saddle point in the PES. This is accepted, since while not stable, this molecule acts as a transition state \cite{ref:H3transition}.
\\

\begin{figure} [H]
\centering
\includegraphics[width=1.0\linewidth]{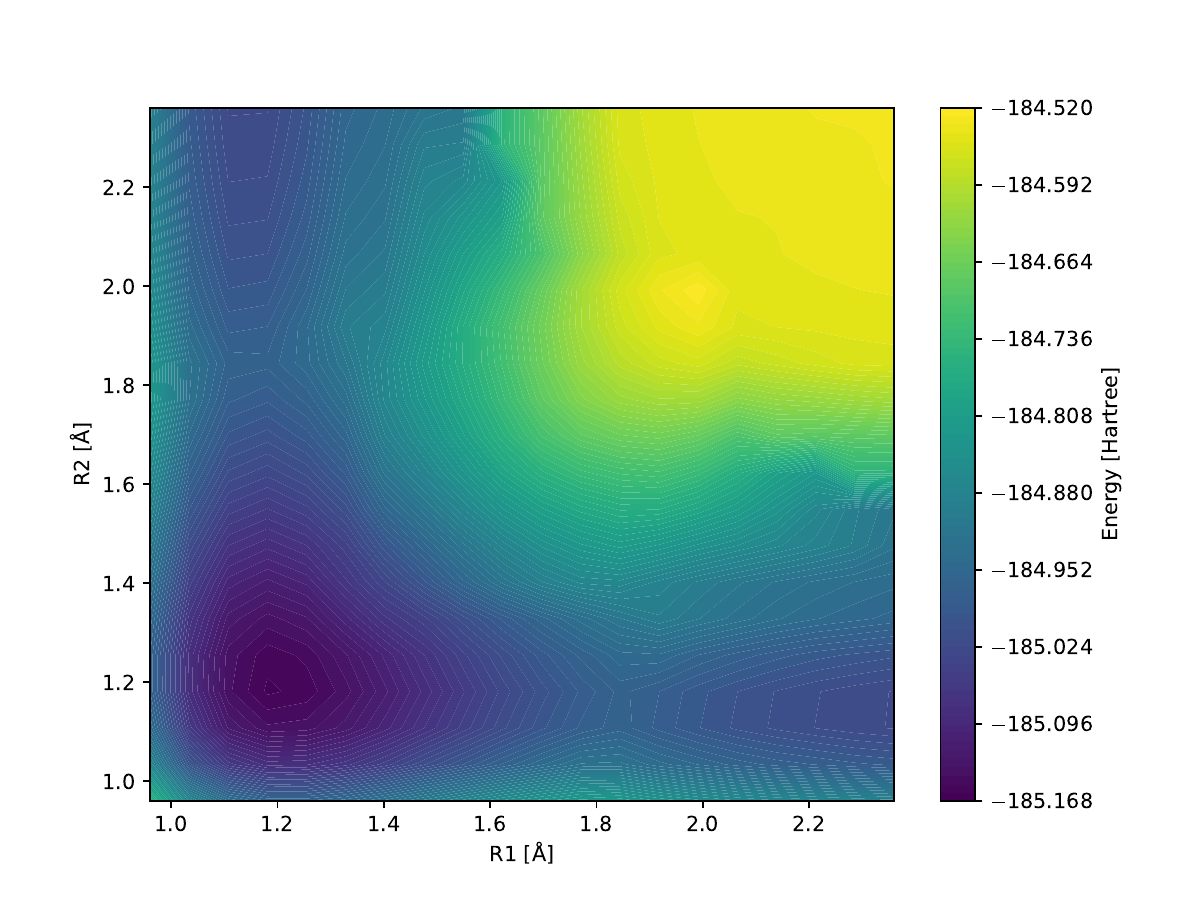}
    \caption{2d PES of co-linear dynamics of two oxygens around a carbon atom.}
    \label{fig:CO2}
\end{figure}

The second PES we plotted is for the dynamics of one carbon atom and two oxygen atoms. We assumed co-linearity for this system as well, with $R_1 [\text{\AA}]$ and $R_2[\text{\AA}]$ the distances between the carbon, to the left and to the right oxygen, respectively. We took $R_1,R_2\in[0.96,2.36]$ which was chosen to capture the interesting regions. We chose AS of 6 electrons and 6 spatial orbitals (12 spin orbitals), which yields a 10-qubit Hamiltonian, whereas the default in PySCF is 22 electrons and 15 spatial orbitals (30 spin orbitals). This means that the lowest-energy $16$ electrons are assumed to simply occupy the eight lowest energy orbitals (16 lowest energy spin-orbitals). The results are shown in \Fig{fig:CO2}

The PES presented in \Fig{fig:CO2} shows that having all the atoms close to each other, is the most preferable state. This result is consistent with the chemical stability of carbon-dioxide under ambient conditions. Other local minima appear when a CO molecule is formed and the other oxygen is far apart, corresponding to the formation of the less (but still) stable molecule of CO. We see that states in which both distances $R_1,R_2$ are too large, are less preferable than the formation of molecules in 
this co-linear settings, as expected.  

\begin{figure} [H]
\centering
\includegraphics[width=1.0\linewidth]{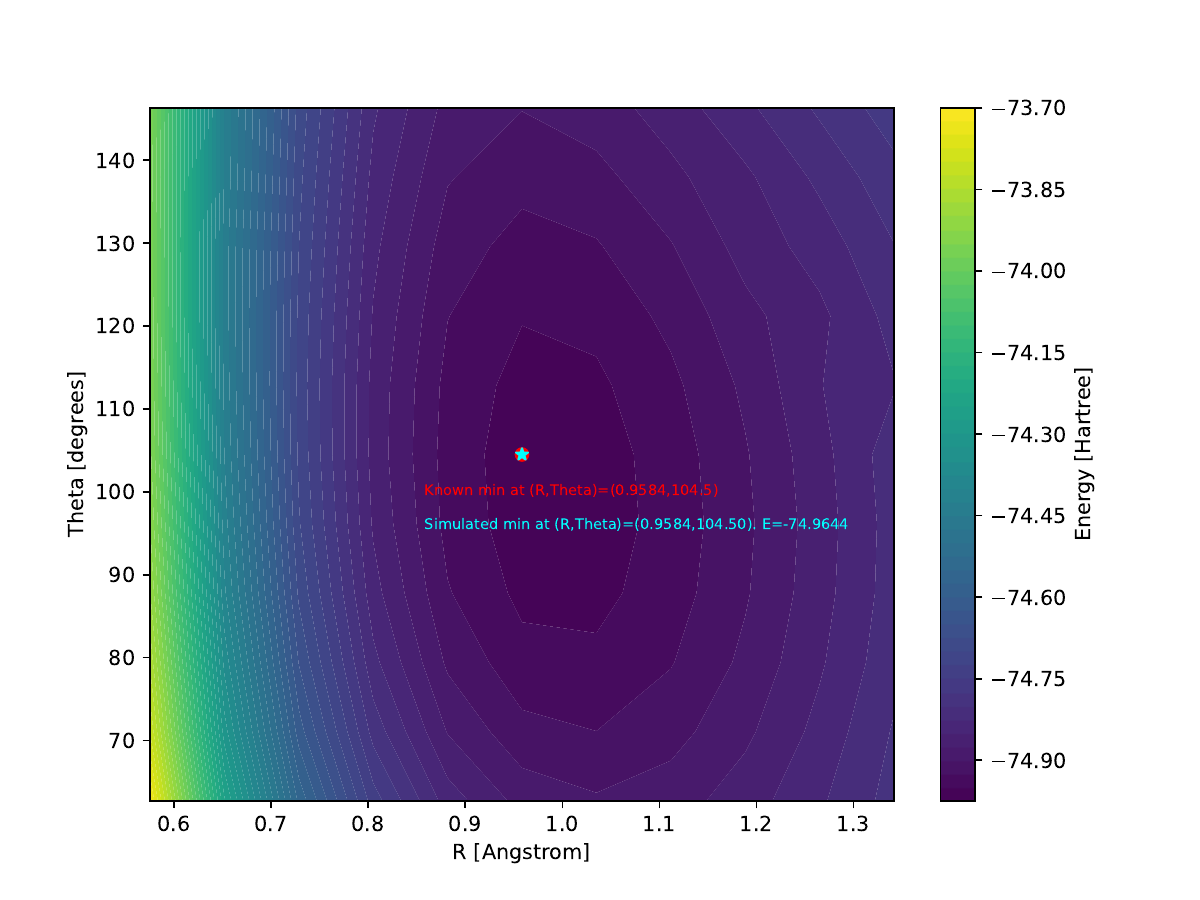}
    \caption{2d PES of water molecule.}
    \label{fig:H2O}
\end{figure}

Finally, we plotted the PES for water, with respect to two parameters: the distance $R[\text{\AA}]$, between each of the hydrogen atoms and the oxygen atom, and the angle $\vec{\theta}[^\circ]$ between the two oxygen-hydrogen bonds. The known parameters at equilibrium are known to be $R_{eq}=0.9582\text{\AA},\vec{\theta}_{eq}=104.5^\circ$. We used a grid of $11\times 11$ points evenly spread (around the equilibrium) in the region $[0.6R_{eq},1.4R_{eq}]\times[0.6\vec{\theta}_{eq},1.4\vec{\theta}_{eq}]$. As expected, we observed minimal energy at the middle point. A minimal active space of 2 electrons and 2 spatial orbitals (4 spin orbitals) was sufficient to demonstrate this behavior. The PES result is shown in \Fig{fig:H2O}.

\subsection{Ground State Energy Simulation of Formic Acid with DQES}
\label{sec:DQES}
\paragraph{Simulation Details}
\label{ref:sim_detailed}
We ran numerous ground-state energy simulations of Formic Acid. We compared the solutions obtained by both standard VQE methods and partial-DQES (see \Sec{sec:alg_rev}) on two qubits. Unless mentioned otherwise, all of the simulations are of Formic Acid (HCOOH) in its lowest energy geometry at absolute zero temperature (see  \App{sec:formic_acid_geometry}). We computed the ground state energy for various levels of approximation, corresponding to different choices of AS. All the Hamiltonians were obtained using PySCF and Qiskit-Nature in the manner discussed in the previous section. Our different choices of active space were: Two electrons and two spatial orbitals (2 qubits), four electrons and four spatial orbitals (6 qubits) and six electrons and six spatial orbitals (10 qubits). The tests were run in parallel on 20 recycled computers for several weeks.

\paragraph{Comparison of Quantum Eigenvalue Algorithms}
For our comparison between VQE and DQES, we measured how the choice of ansatz circuit affects both algorithms. Most work on VQE is done with some choice of an ansatz and an optimizer,
while the initial state is simply $\ket{000 \ldots 00}$ for all
qubits. Of course there are a few exceptions (e.g. a piece of work
named subset-VQE, and various additional works); we intend to deal
with those in the future, and compare our method to the main ones. 

We have seen that the use of MUBs could be extremely important and
even crucial: E.g., in the case of a singlet and a degenerate triplet,
one could accidentally use an ansatz that will solely be restricted to
the huge barren plateau. MUB initialization prevents that.

Let us now decide how to compare to the above case in which the initial
state is always $\ket{000 \ldots 00}$. In all such papers, the
flexibility is with choosing a vector $\vec{\theta}$
of all angles in the ansatz, commonly a random choice.  If the
optimizer does not lead to a convergence,
another attempt or a few attempts are then done.
Commonly there
is no deep understanding of how many attempts should be applied, so we
can for example compare to just five attempts (see below).

However, the intuition of spanning the entire quantum space is what
causes us (and of course~\cite{ref:AMM2024}) to use MUBs. So the full DQES,
for
small problems, used $2^n \times (2^n + 1)$ states. It makes sense to
compare to an identical number of random attempts, if the ansatz is
expressive (can reach any point in the quantum space). Commonly the
ansatz is not expressive unless in papers dealing with very few (e.g.,
2-3) qubits. So then, some measure of ``nicely spread'' states can
maybe be defined.  E.g., the Haar measure is sometimes used, but
commonly only for the full quantum space. We are aware of very few
papers that try many
random initial states, e.g \cite{ref:Miyahara} and \cite{ref:KimOz2022}. In the future we intend to check how many such
random states, so we can try to use an identical number and compare. We also plan to compare the performance of our method with other clever randomization techniques, for example, see \cite{ref:Grant2019} , in which only a subset of the parameters is randomly initialized.

Given the above information, one fair way of comparing is to check
paper by paper and use a small number of random choices as done there
and compare to our attempt with larger number of initial states, that
commonly depends on some deterministic yet arbitrary choice of
partial efficient DQES (\cite{ref:AMM2024}) or the full DQES if the system is
sufficiently small.

Alternatively, we can adjust the number of random vectors $\vec{\theta}$, to
be exactly identical to the number of our initial states that uses
full or partial DQES.

Theoretically, For a fully
expressive ansatz and full DQES we do not know yet what will be
better, to use all $2^n \times (2^n + 1)$ states or to use
$2^n \times (2^n + 1)$ random $\vec{\theta}$ vectors. We believe that both
methods will be great in different ways, for analyzing BPs. However,
this theoretical comparison is far beyond the scope of our program
here. It could take years or decades till such a theoretical
comparison will be performed.

Here, for non-expressive ansatz, versus some deterministic yet
arbitrary partial DQES, we can only compare
heuristics.  This is done here, see below.  We plan to also compare,
for very small cases, full DQES with fully expressive ansatz, and with
the same number of attempts.

Here, we tested two ansatze:  \texttt{UCCSD}~\cite{ref:UCCSD} and \texttt{EfficientSU2}~\cite{ref:EfficientSU2}. Their circuit representation is given in \App{sec:circuits}. When using partial-DQES, we optimize once over all MUB states on two qubits with the other qubits fixed to $\ket 0$. We chose the initial parameters of the ansatz all to zero $(\vec{\theta}_0=\vec 0)$. Hence, this algorithm is completely deterministic yet arbitrary. In contrast, VQE is initialized with random initial parameters for each trial. We measured the effect of a constant number of VQE trials (5 trials as a baseline) and an adaptive number of VQE trials, which is set to be the same number of trials as the amount of MUB states for DQES, denoted by $N$. Specifically, for 2 qubits $N=19$, for 6 qubits $N=286$ and for 10 qubits $N=856$\footnote{After removing redundant $\ket0$ states}. We classically found the ground-state energy $E_0$ and evaluated the \textit{ground-state energy error} of each trial, defined as $\Delta E=E-E_0$ where $E$ is the optimized energy from each trial. The results are summarized in Table \ref{tab:main_results}. For $\mu=1.4\times 10^{-3}$ [Hartree] known as \textit{chemical accuracy} we also checked if the results are precise, namely if $\Delta E <\mu $. This is a well-known standard of precision in chemistry \cite{ref:pople1999}.

\begin{table} [H]
\begin{tabular}{@{} l lll} 
\toprule
\multicolumn{1}{c}{Test}  &  Method& $\overline{\Delta E}$ [Hartree] $\downarrow$&$min[\Delta E]$ [Hartree] $\downarrow$\\
\midrule
\multirow{3}{*}{2 qubits - \texttt{UCCSD}}& VQE - 20 trials& $\mathbf{8.5\times 10^{-15}}$&$\mathbf{<10^{-16}}$\\
 & VQE - 5 trials& $1.1\times 10^{-14}$&$2.7\times10^{-15}$\\
 & DQES (ours)& $6.4\times 10^{-2}$&$4.4\times 10^{-16}$\\
 \bottomrule
\multirow{3}{*}{6 qubits - \texttt{UCCSD}}& VQE - 286 trials& $\mathbf{1.4\times10^0}$&$1.3\times 10^0$\\
 & VQE - 5 trials& $\mathbf{1.4\times10^0}$&$1.4\times 10^0$\\
 & DQES (ours)& $1.7\times10^0$&$\mathbf{3.9\times 10^{-5}}$\\
\bottomrule
 \multirow{3}{*}{10 qubits - \texttt{UCCSD}}& VQE - 856 trials& $4.3\times10^0$&$4.2\times 10^0$\\
 &VQE - 5 trials&$4.3\times10^0$&$4.2\times 10^0$\\
 & DQES (ours)& $\mathbf{2.2\times10^0}$&$\mathbf{6.4\times 10^{-3}}$\\
 \bottomrule
 \multirow{3}{*}{2 qubits - \texttt{EfficientSU2}}& VQE - 19 trials& $1.1\times 10^{-14}$&$\mathbf{<10^{-16}}$\\
 & VQE - 5 trials& $\mathbf{8.9\times10^{-15}}$&$3.3\times10^{-15}$\\
 & DQES (ours)& $3.9\times10^{-9}$&$1.3\times10^{-15}$\\
 \bottomrule
 \multirow{3}{*}{6 qubits - \texttt{EfficientSU2}}& VQE - 286 trials& $3.3\times 10^{-2}$&$1.62\times 10^{-3}$\\
 & VQE - 5 trials& $\mathbf{2.6\times 10^{-2}}$&$2.32\times 10^{-2}$\\
 & DQES (ours)& $5.1\times 10^{-2}$&$\mathbf{1.58\times 10^{-3}}$\\
 \bottomrule
 \multirow{3}{*}{10 qubits - \texttt{EfficientSU2}}& VQE - 856 trials& $2.7\times 10^{-1}$&$3.0\times 10^{-2}$\\
 & VQE - 5 trials& $\mathbf{2.0\times 10^{-1}}$&$7.6\times 10^{-2}$\\
 & DQES (ours)& $4.4\times 10^{-1}$&$\mathbf{1.8\times 10^{-2}}$
\end{tabular}
 \caption{Comparison of DQES and VQE optimization results on the qubit Hamiltonian of Formic Acid.}
 \label{tab:main_results}
\end{table}
As seen in Table \ref{tab:main_results}, when the number of qubits and the depth of the ansatz increase, the minimum energy gap for VQE becomes significant ($\sim$ 1 [Hartree] and higher). In contrast, DQES manages to achieve results within chemical accuracy. Note that the average Energy Error in VQE is usually lower than in DQES.

\begin{figure} [H]
\begin{subfigure}[b]{0.5\textwidth}
    \centering
    \includegraphics[width=0.99\linewidth]{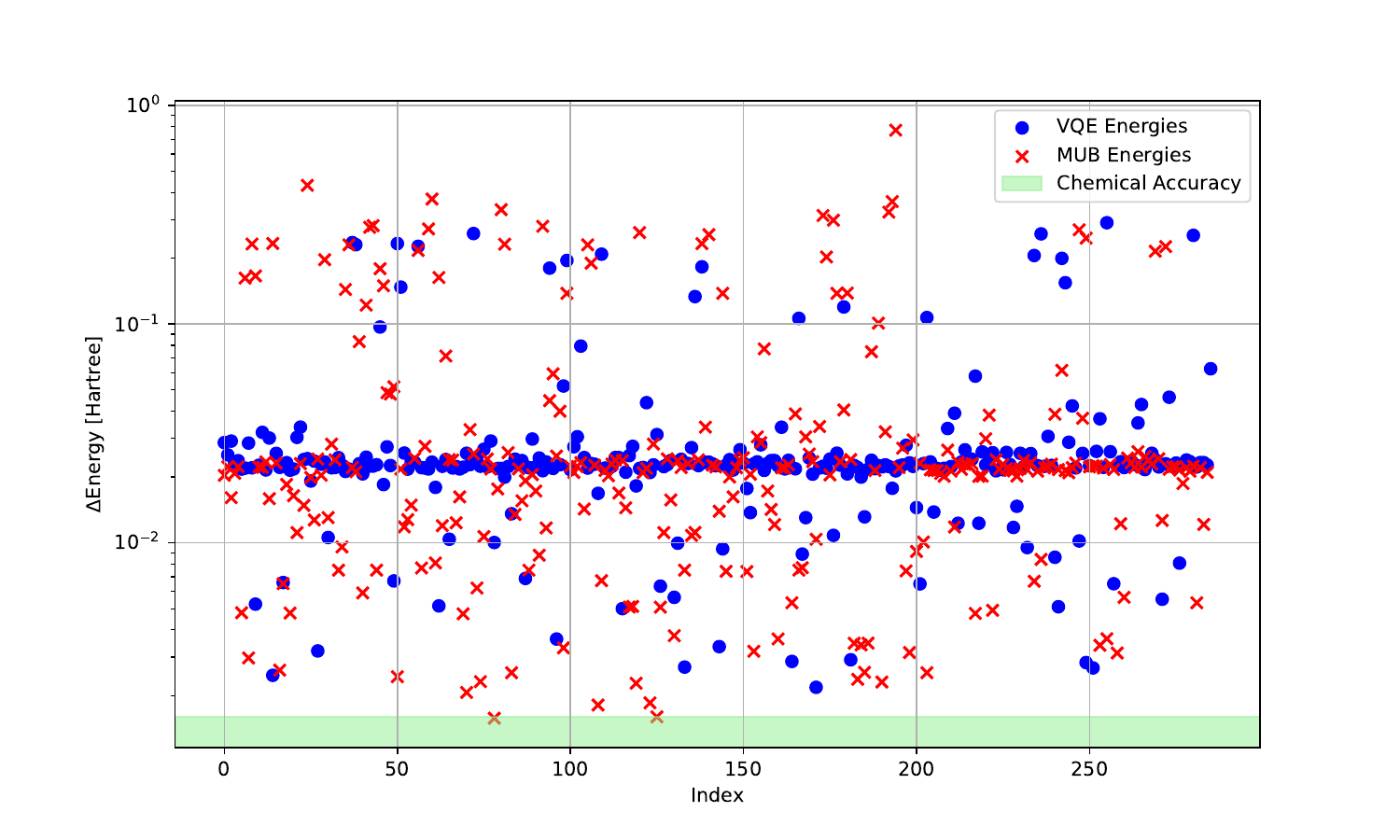}
\end{subfigure}
\begin{subfigure}[b]{0.5\textwidth}
    \centering
    \includegraphics[width=0.99\linewidth]{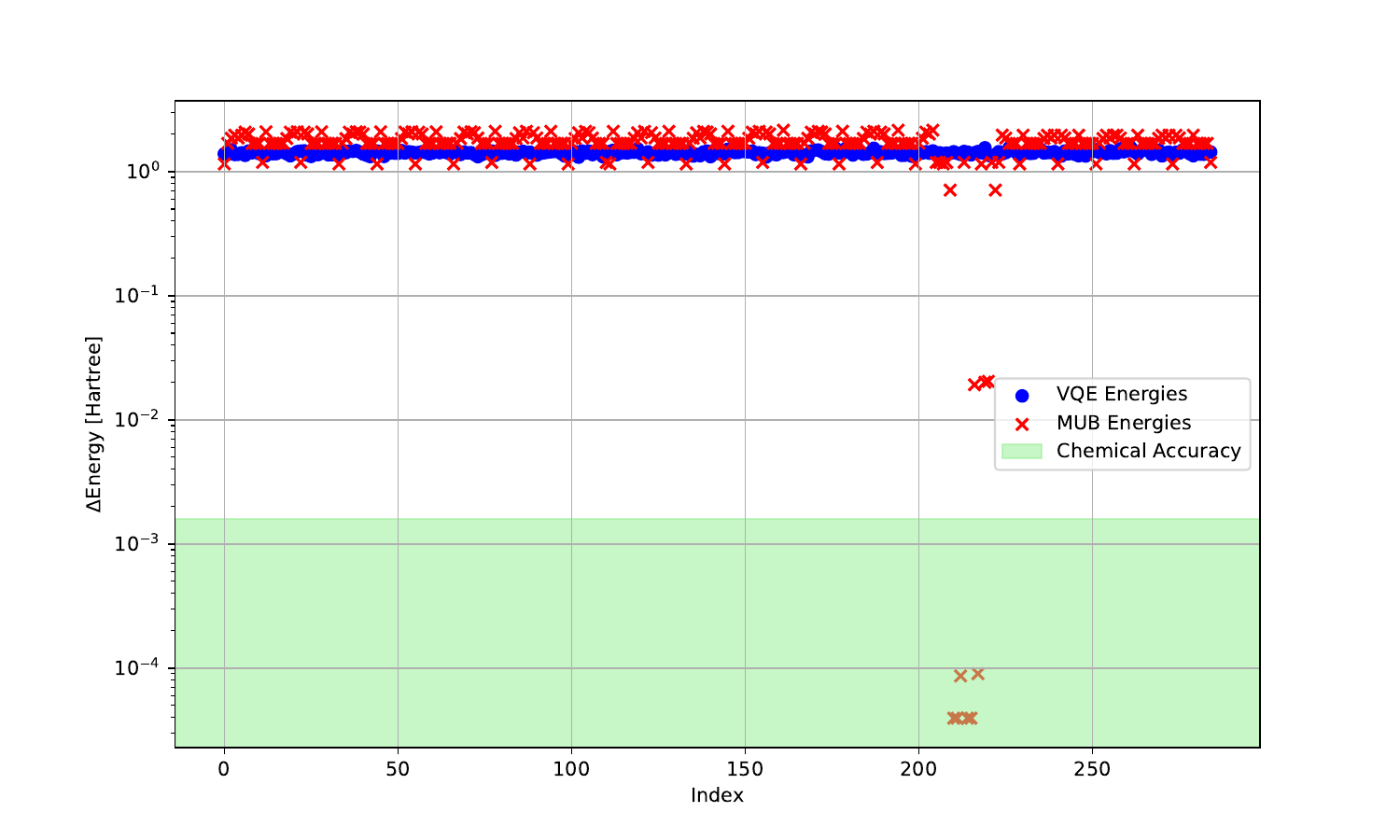}
\end{subfigure}
    \caption{The Energy Error spread in 6 qubits \texttt{EfficientSU2} (left; in log-scale) and \texttt{UCCSD} (right) in VQE trials (Blue) and MUB trials (Red).}
    \label{fig:energy_comparison_6_qubit}
\end{figure}

\begin{figure} [H]
\begin{subfigure}[b]{0.5\textwidth}
    \centering
    \includegraphics[width=0.99\linewidth]{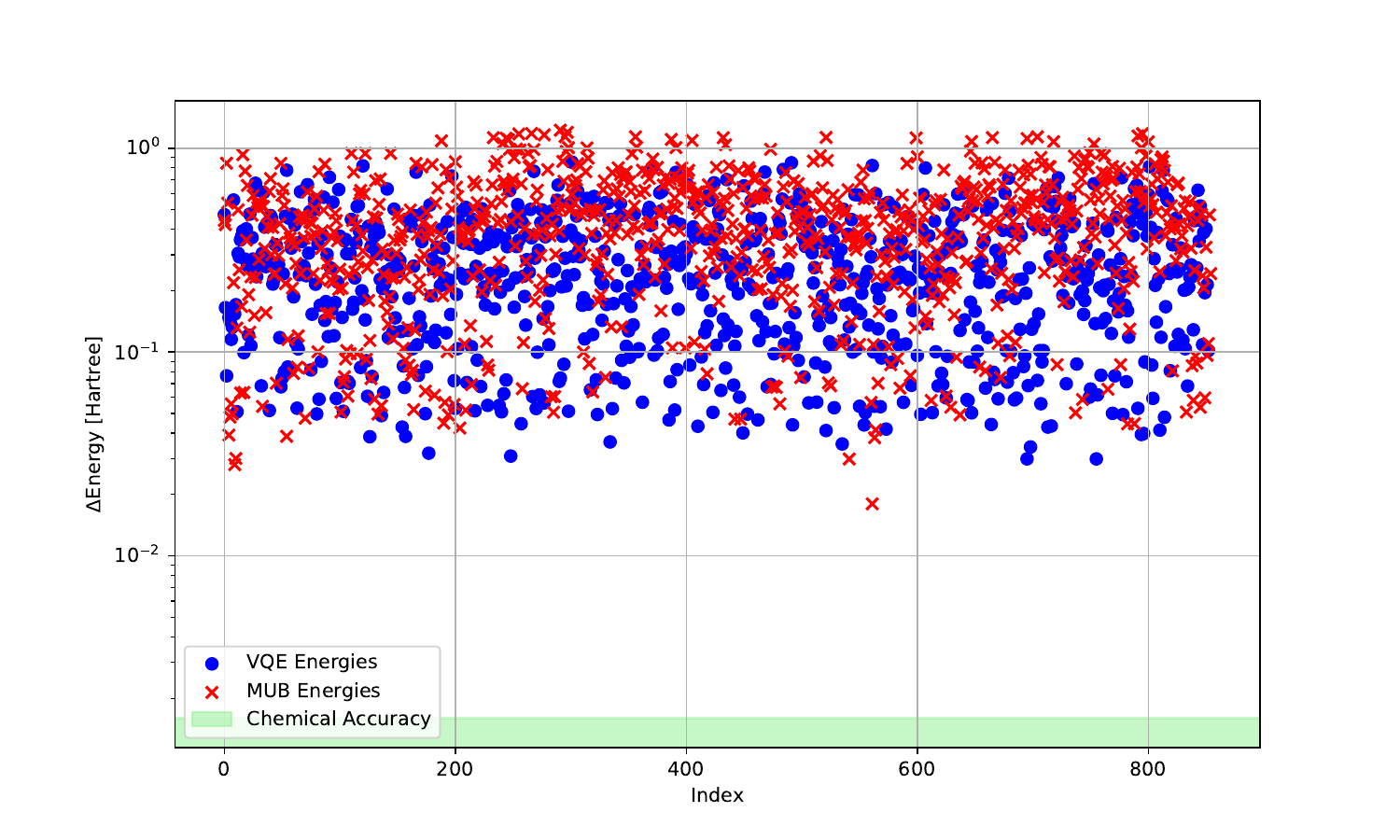}
\end{subfigure}
\begin{subfigure}[b]{0.5\textwidth}
    \centering
    \includegraphics[width=0.99\linewidth]{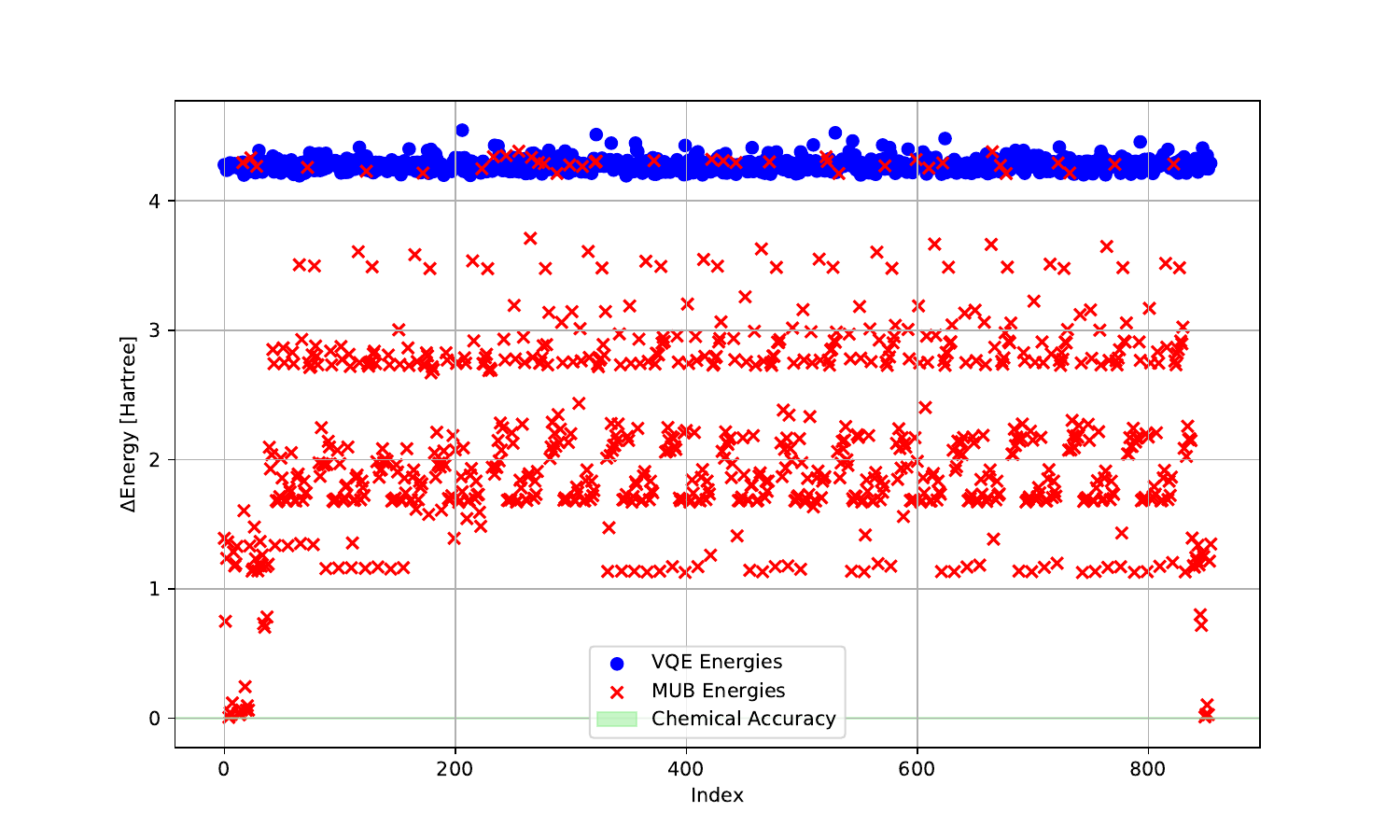}
\end{subfigure}
    \caption{Energy Error spread in 10 qubits \texttt{EfficientSU2} (left) and \texttt{UCCSD} in log (right) in VQE trials (Blue) and MUB trials (Red).}
    \label{fig:energy_comparison_10_qubit}
\end{figure}

We collected the final energy results from all trials of both algorithms. The results are summarized in \Fig{fig:energy_comparison_6_qubit} for the 6-qubit test on both ansatze, and in \Fig{fig:energy_comparison_10_qubit} on the 10-qubit test.

According to \Fig{fig:energy_comparison_6_qubit}, in both ansatze the minimum of DQES trials is well below the minimum of VQE trials. When using the \texttt{EfficientSU2} ansatz, two MUB trials reached chemical accuracy, while all VQE trials did not. When using \texttt{UCCSD} ansatz, six MUB trials reached chemical accuracy. According to \Fig{fig:energy_comparison_10_qubit}, for both ansatze, all VQE trials got stuck in a significant energy gap, while DQES trials were scattered in several regions. In the \texttt{UCCSD} case, some of the DQES trials even reached low energy gap (yet not in chemical accuracy range, see \Fig{fig:energy_comparison_6_UCCSD_log_scale} in  \App{sec:logplot}). From these two figures, we find that \texttt{EfficientSU2} tends to yield a high variance in optimized energy, while \texttt{UCCSD} yields either significantly high energy gaps or extremely low energy gaps.\\

\begin{figure} [H]
\begin{subfigure}[b]{0.5\textwidth}
    \centering
    \includegraphics[width=0.99\linewidth]{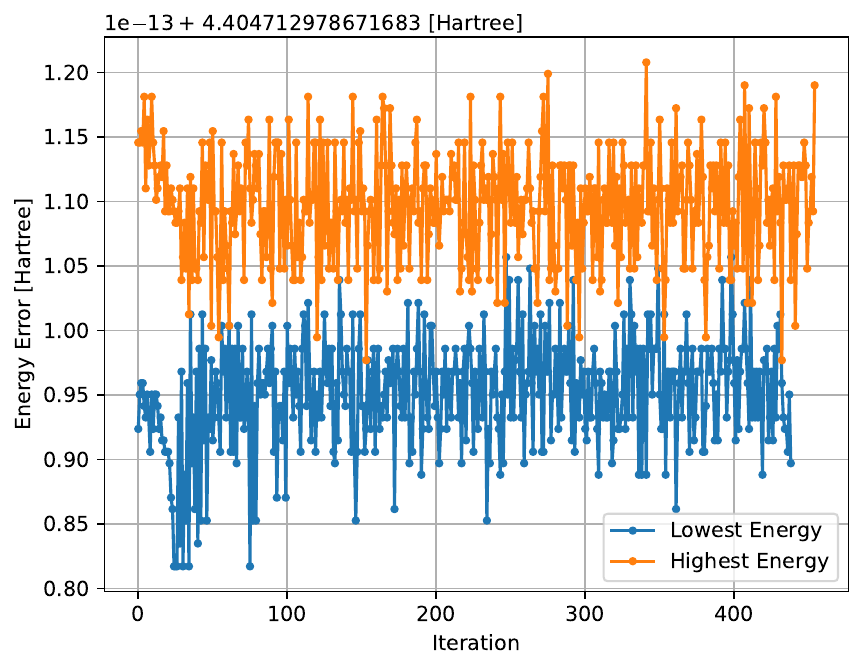}
\end{subfigure}
\begin{subfigure}[b]{0.5\textwidth}
    \centering
    \includegraphics[width=0.99\linewidth]{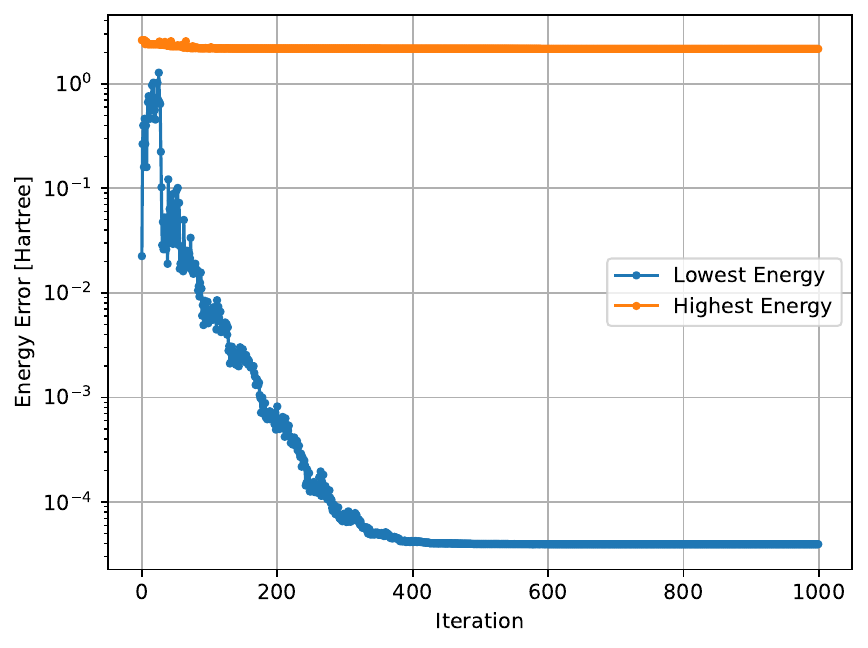}
\end{subfigure}
\caption{
    Optimization plots of the 6 qubit \texttt{UCCSD} task with the lowest and highest energy trials out of all the VQE attempts (left) and MUB attempts (right).
    }
\label{fig:optimization_plot_formic_4_UCCSD}
\end{figure}

\begin{figure} [H]
\begin{subfigure}[b]{0.5\textwidth}
    \centering
    \includegraphics[width=0.99\linewidth]{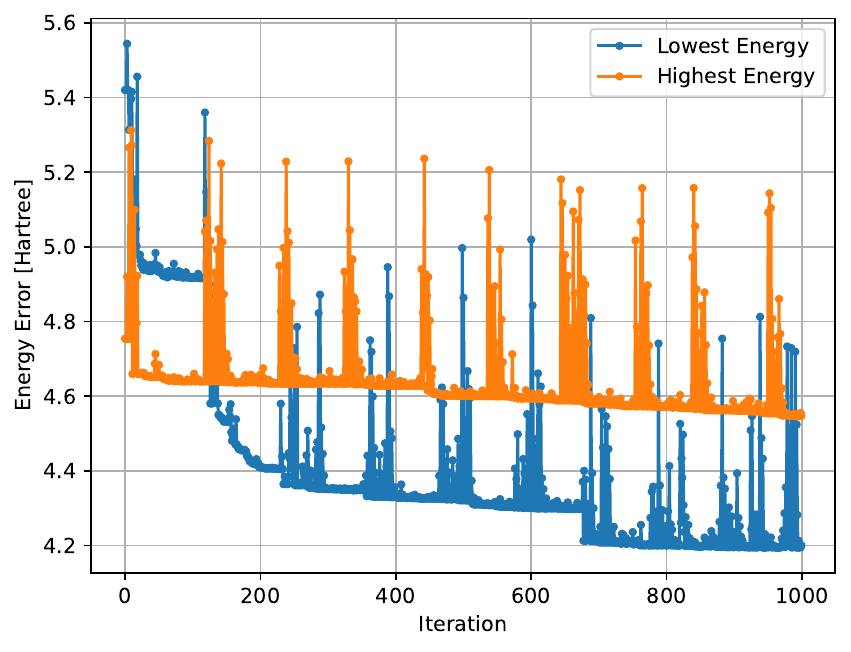}
\end{subfigure}
\begin{subfigure}[b]{0.5\textwidth}
    \centering
    \includegraphics[width=0.99\linewidth]{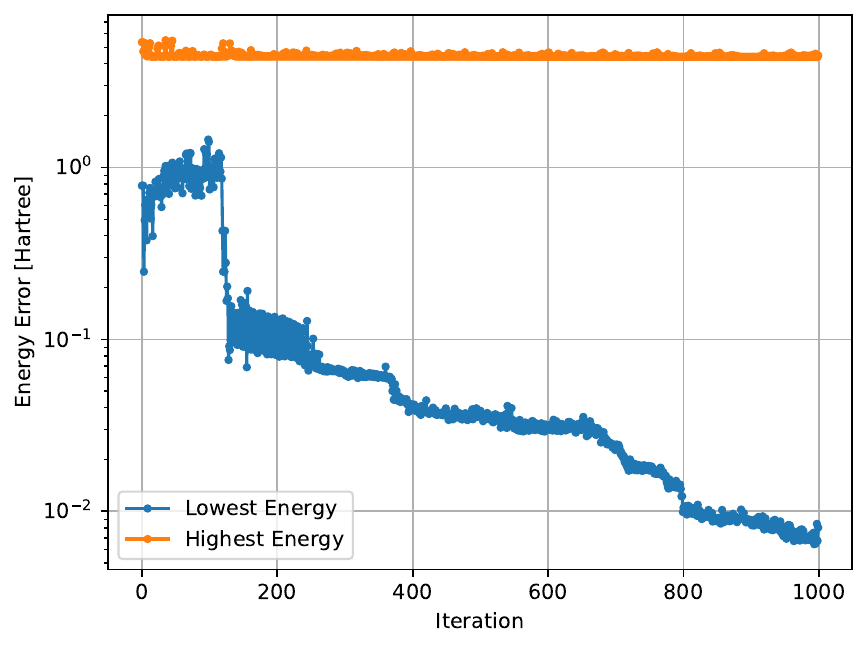}
\end{subfigure}
    \caption{Optimization plots of the 10 qubit \texttt{UCCSD} task with the lowest and highest energy trials out of all the VQE attempts (left) and MUB attempts (right). }
    \label{fig:optimization_plot_formic_6_UCCSD}
\end{figure}

Next, for each method, we present the best and worst convergence performances. Results for the 6-qubit case and the 10-qubit case (both with the \texttt{UCCSD} ansatz) are presented in \Fig{fig:optimization_plot_formic_4_UCCSD} and \Fig{fig:optimization_plot_formic_6_UCCSD}, respectively.

In the 6 qubit test, summarized in \Fig{fig:optimization_plot_formic_4_UCCSD}, in the VQE case both the best and the worst trials are similar and zigzag around a constant and significant energy gap ($\sim$ 4 Hartree), showing that this optimization in this landscape is ineffective. In contrast, DQES in the best case optimizes normally and even achieves convergence at the 400th iteration. It is interesting to note that in the worst case it appears to get stuck on a high energy gap, similar to VQE.
In the 10 qubit test, as seen in \Fig{fig:optimization_plot_formic_6_UCCSD}, VQE still appears to get stuck on different but significant energy gaps ($\sim$ 4 Hartree). In DQES, the worst case does seem to get stuck on a high energy gap like before. While the best case optimizes normally, it doesn't reach absolute convergence in 1000 iterations, though it outperforms the best VQE trial.
\paragraph{Further variations}
As a final comparison, we test the effect of the initial state of the ansatz circuit to the optimization landscape. For each VQE trial, we randomize the initial state $\ket{\psi_0}$ to be a random state in the computational basis. For each DQES trial, we first select the qubit pair to have the MUB state, and then the \textit{rest} of the qubits are randomly sampled to be a state in the computational basis. The results for these tests are in Table \ref{tab:variations}.
\begin{table} [H]
\begin{tabular}{@{} l lll}
\toprule
\multicolumn{1}{c}{Test}  &  Method& $\overline{\Delta E}$ [Hartree] $\downarrow$&$min[\Delta E]$ [Hartree] $\downarrow$\\
\midrule
\multirow{2}{*}{6 qubits - \texttt{UCCSD}}& VQE + random $\ket {\psi_0}$ - 286 trials& $\mathbf{5.6\times 10^{-1}}$&$2.0\times 10^{-2}$\\
 & DQES + random $\ket {\psi_0}$& $6.1\times 10^{-1}$&$\mathbf{3.9\times 10^{-5}}$\\
 \bottomrule
\multirow{2}{*}{6 qubits - \texttt{SU2}}& VQE + random $\ket {\psi_0}$ - 286 trials& $\mathbf{3.7\times 10^{-2}}$&$2.0\times 10^{-3}$\\
 & DQES + random $\ket {\psi_0}$& $7.2\times 10^{-2}$&$\mathbf{1.8\times 10^{-3}}$\\   
\end{tabular}
 \caption{Summary of results of the random initialization state on the 6 qubit problems. }
 \label{tab:variations}
\end{table}
According to Table \ref{tab:variations}, we observe an improvement in the minimum energy gap for both algorithms compared to the results seen in Table \ref{tab:main_results}. Even then, the energy gap in the \texttt{UCCSD} test is significantly higher in VQE than in DQES, while in \texttt{EfficientSU2} it is less significant.
\subsection{Transition state optimization}
\label{sec:TS_result}
Throughout the following experiments we used the computational chemistry library ORCA \cite{ref:orca} which has both NEB and IRC algorithms implemented. As mentioned in \Sec{sec:IRC} we first want to find the transition state and then utilize it to find the full reaction coordinate.
\begin{figure}[H]
    \centering
    \includegraphics[width=0.4\linewidth]{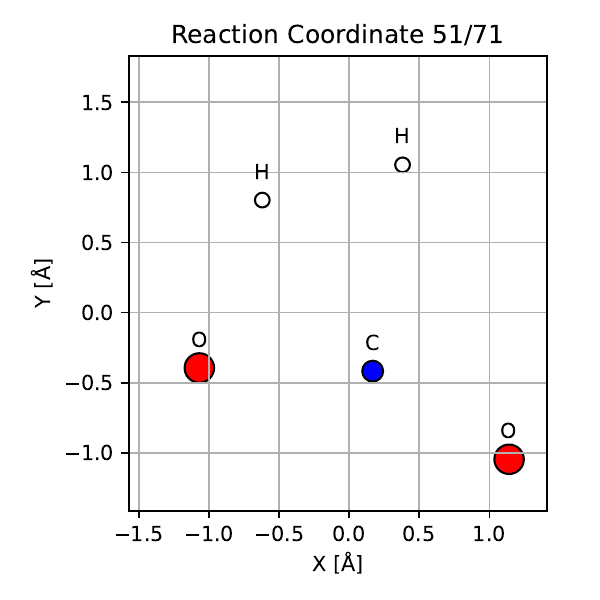}
    \caption{The Transition State as calculated by the NEB-TS optimization method.}
    \label{fig:transition_state}
\end{figure}
According to \Fig{fig:transition_state}, the transition state happens when all the atoms H-O-C-H have electrons connecting them, yielding a highly-unstable and high-energy configuration. We expect the states before it to have the H$_2$ molecule further away and the states after it to have the H-H bond broken, with each hydrogen going to the oxygen or to the carbon atoms. The energy of this configuration is $-189.629$, [Hartree] while the energy of the reactants (H$_2$+CO$_2$) is $-189.738$ [Hartree] and that of the product (HCOOH) is $-189.735$ [Hartree]. These values tell us that the reaction has activation energy $\Delta E_a=0.109$ [Hartree], and that it is endothermic with an energy difference of $0.003$ [Hartree].
\subsection{IRC calculation}
\begin{figure}[H]
    \centering
    \includegraphics[width=0.6\linewidth]{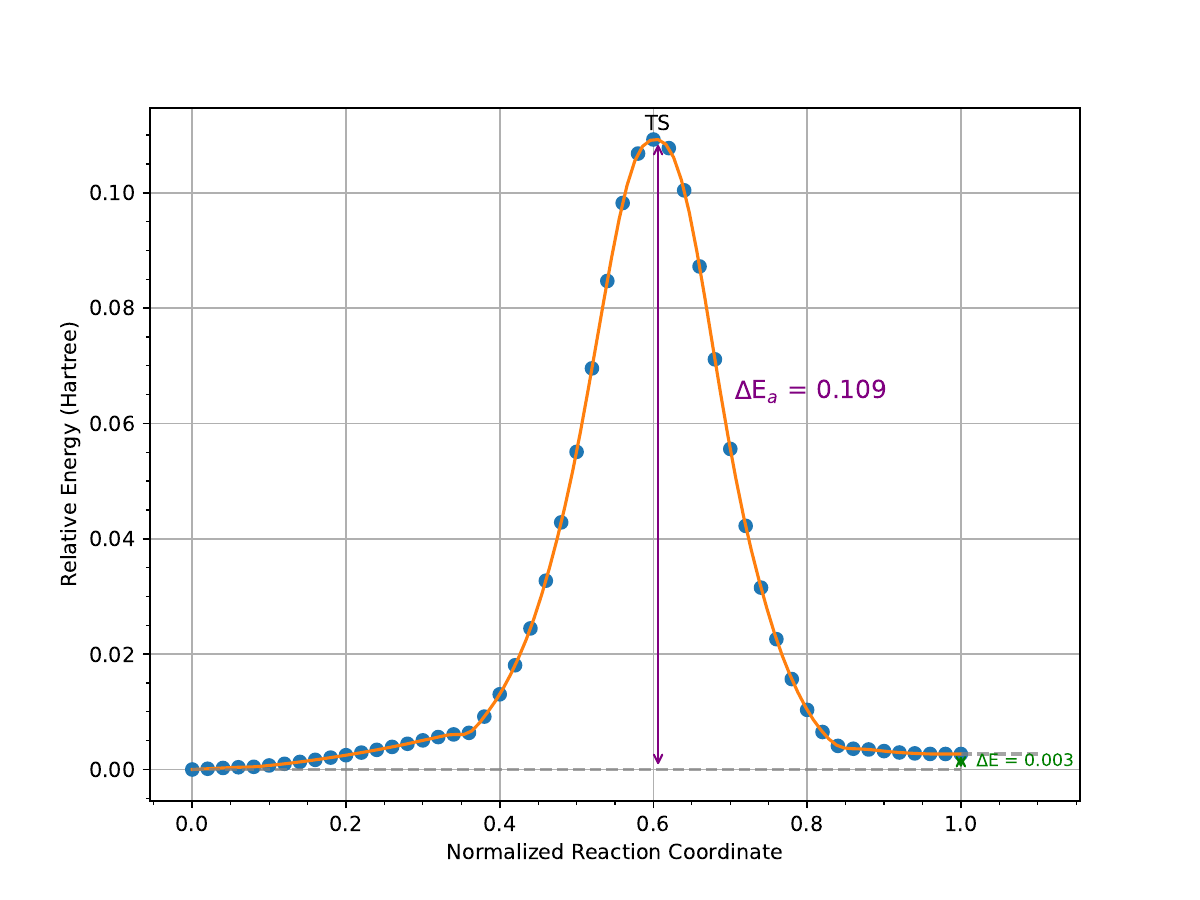}
    \caption{Intrinsic Reaction Coordinate Plot of the formic acid reaction obtained by the IRC method on the transition state from \Fig{fig:transition_state}}
    \label{fig:IRC_plot}
\end{figure}

We wish to find the full reaction coordinate of the reaction. The results of the IRC method on the previously obtained transition state is given in \Fig{fig:IRC_plot}.
According to \Fig{fig:IRC_plot}, the entire reaction has one transition state, flowing from the reactants to the product, with no other intermediates states. The left side of the graph requires many iterations since it includes the path the hydrogen takes to come close to the carbon dioxide. 

\begin{figure}[H]
     \centering
     \begin{subfigure}[b]{0.24\textwidth}
         \centering
         \includegraphics[width=0.99\linewidth]{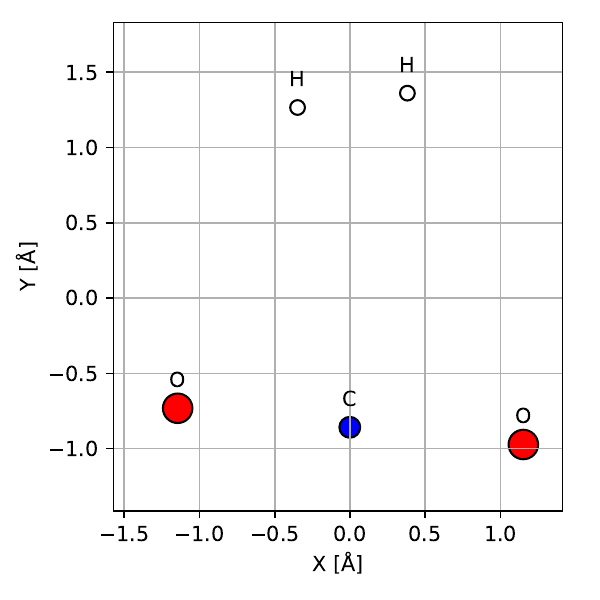}
         \label{fig:frame_038}
         \caption{}
     \end{subfigure}
     \begin{subfigure}[b]{0.24\textwidth}
         \centering
         \includegraphics[width=0.99\linewidth]{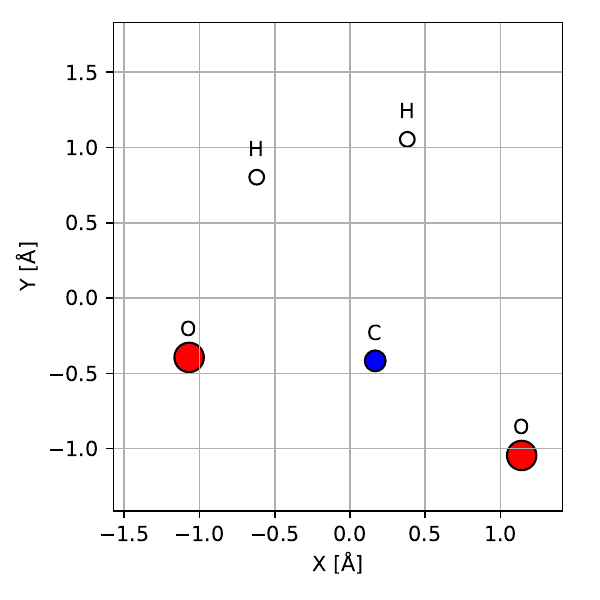}
         \label{fig:frame_050}
         \caption{}
     \end{subfigure}
     \begin{subfigure}[b]{0.24\textwidth}
         \centering
         \includegraphics[width=0.99\linewidth]{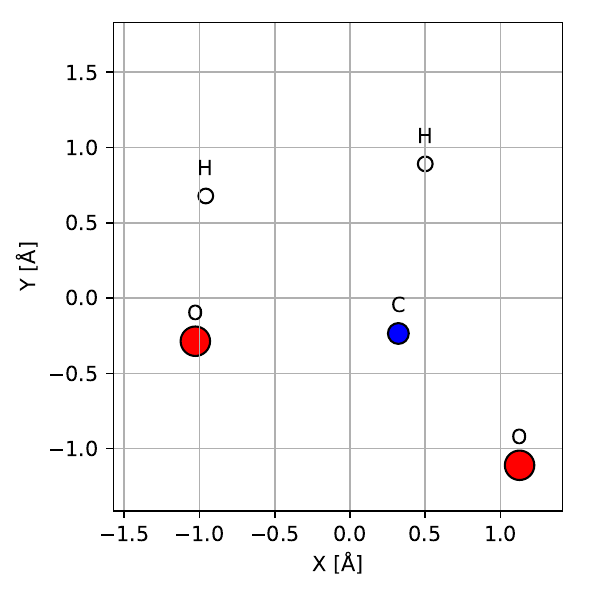}
         \label{fig:frame_057}
         \caption{}
     \end{subfigure}
     \begin{subfigure}[b]{0.24\textwidth}
         \centering
         \includegraphics[width=0.99\linewidth]{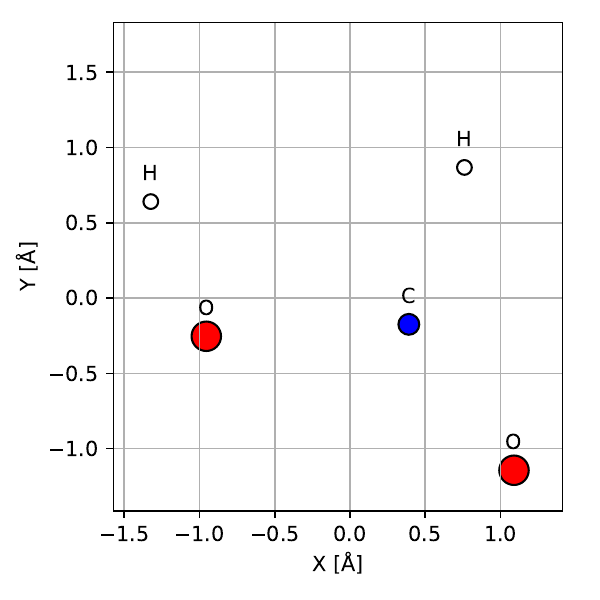}
         \label{fig:frame_070}
         \caption{}
     \end{subfigure}
     
        \caption{Four different snapshots of the molecule configurations on the reaction coordinate. Snapshot (a) is the initial configuration, where H$_2$ is separated from CO$_2$. It is slightly tilted to break symmetry. Snapshot (b) is the transition state as was found in \Sec{sec:TS_result}. Snapshot (c) shows the state where the H-H bond has broken and head hydrogen forms its own bond with oxygen \& carbon. Finally, (d) shows the product, formic acid, which is reached after the hydrogen atoms move apart due to steric repulsion.}
        \label{fig:reaction_summary}
\end{figure}

To further inspect the reaction, we inspect a few selected snapshots of the reaction throughout the optimized reaction coordinate, summarized in \Fig{fig:reaction_summary}.

\section{Discussion and Future Research}
\label{sec:discussion}
In this work, we studied formic acid from a hybrid quantum-classical point of view. The reaction we studied has connection to carbon fixation and origin of life as was discussed in \Sec{sec:intro}. Our analysis is an initial step in a research about these connections. Applying the tools we used in this work such DQES with VQE, PES calculation, as well as NEB and IRC algorithms, on more complex set-ups that involve catalysts, may shed more light on the significance of the process, and hopefully lead to interesting applications.  

Our PES calculations for the two co-linear cases demonstrate known phenomena using a quantum-inspired tool and represent an important step before moving on to the more complicated optimization process of finding the reaction coordinate. We consider this a method for future work that can be run on both classical and quantum computers, with improved running time due to symmetries that can be exploited in the qubit formulation. Exploring systems with more degrees of freedom remains a challenge we wish to address in the future by using coarse-graining methods and a clever choice of coordinates to ignore due to symmetries. We suggest also studying the following alternative reaction resulting in formic acid 
\[
\text{CO}+\text{H}_2\text{O}\rightarrow\text{HCOOH},
\]
with the same methods applied in this work. As a first step, we already presented the PES of water, by using our quantum-inspired analysis (\Sec{sec:PESRes}).
Our results for VQE with Mutually Unbiased Bases provide evidence for the success of DQES as a discrete search that avoids barren plateaus and local minima where regular VQE fails. The results, summarized in \Tab{tab:main_results}, show that in small problems (in this case, AS of two electrons and two spatial orbitals), both VQE and DQES perform well, with minimal error. However, as the active space expands to 4 electrons, 4 spatial orbitals and 6 electrons, 6 spatial orbitals, standard VQE fails to converge reliably, likely due to barren plateaus~\cite{ref:mcclean2018}. DQES, in contrast, finds at least one solution within chemical accuracy, supporting the hypothesis that structured ansatze avoid flat landscapes\cite{ref:oz2022}. This is consistent with recent Hessian-based analysis showing improved trainability in low-curvature regions~\cite{ref:dauphin2021}.

We note that \textit{on average}, DQES performs worse than VQE, which aligns with intuitions of a wide search in various regions both good and bad. This lets the minimum energy gap in DQES be significantly lower than in VQE, which implies random VQE initializations are not a successful search strategy. This is further illustrated in \Fig{fig:energy_comparison_6_qubit} and \Fig{fig:energy_comparison_10_qubit} which show that most of the trials in both DQES and VQE 'get stuck' on higher-energy barren plateaus or local minima, while only a select few MUB states manage to slip by and find the global minimum, up to chemical accuracy.

Another behavior that we found is that it is harder to optimize with \texttt{UCCSD} compared to \texttt{EfficientSU2}, while the closest results to the global minimum are when optimizing \texttt{UCCSD}. We believe this is because of the large depth of the \texttt{UCCSD} ansatz that makes it more likely to exhibit barren plateaus. This is illustrated best in \Fig{fig:optimization_plot_formic_4_UCCSD} and \Fig{fig:optimization_plot_formic_6_UCCSD} where it is clear that VQE struggles to optimize due to barren plateaus or local minima, while DQES manages to find the initial states to avoid them.

The results of the modified algorithms in \Tab{tab:variations} show improvement in the results for both algorithms, indicating that the decision of $\ket{\psi_0}=\ket 0$ is arbitrary. Even in these variations, DQES still outperforms VQE in the best case, which means it is a better exhaustive search than simply selecting a random computational basis state as the initial state. These results provide intuition to expect that full-MUB DQES will provide even better results.

Future directions include exploring quantum subspace techniques like CSVQE~\cite{ref:vanbeeumen2024}, which have been shown to extract better energy estimates from mid-circuit states, even when optimization converges to local minima. Additionally, landscape diagnostics using Hessian eigenvalue spectra~\cite{ref:oz2022} or Fisher information may be applied to characterize trainability zones in high-dimensional ansatze. As was suggested in \Sec{sec:DQES}, a comparison with the alternative initialization technique of Grant et al.~\cite{ref:Grant2019}, and an informed choice for ideal number of random runs (according to \cite{ref:Miyahara,ref:KimOz2022}), are valuable directions to us.  

We plan to extend our approach to larger, more complex catalytic models, including metal-organic frameworks and heterogeneous catalytic systems, validating our methodology on available quantum hardware platforms.
%
%
\section*{Acknowledgments}
We would like to clarify that a portion of this work, specifically the implementation of the DQES method, was conducted as part of the initial phase of the XPRIZE Quantum Applications competition, utilizing the framework described in Ref.~\cite{ref:AMM2024}. This specific contribution was performed without financial support. We thank Michael Iv, Linoy Nagar Shaul, Thales Silva and Uri Peskin for fruitful discussions. T.M and O.G thank the
Quantum Computing Consortium of Israel Innovation Authority for financial
support. This research project was partially supported by the Helen Diller
Quantum Center at the Technion.
%
%
\appendix
\newpage
\usetikzlibrary{shapes.geometric, arrows.meta, positioning, calc, shadows}

\tikzstyle{process} = [rectangle, minimum width=3.5cm, minimum height=1cm, text centered, draw=black, fill=blue!10, rounded corners, drop shadow]
\tikzstyle{qiskit} = [rectangle, minimum width=3.5cm, minimum height=1cm, text centered, draw=black, fill=purple!10, rounded corners, drop shadow]
\tikzstyle{decision} = [diamond, minimum width=3cm, minimum height=1cm, text centered, draw=black, fill=green!10, drop shadow]
\tikzstyle{io} = [trapezium, trapezium left angle=70, trapezium right angle=110, minimum width=3cm, minimum height=1cm, text centered, draw=black, fill=orange!10, drop shadow]
\tikzstyle{arrow} = [thick,->,>=Stealth]
\tikzstyle{line} = [thick,-]
\section{Hamiltonians from VQE and PES experiments}
\label{sec:Hams}
Throughout this work, we ran VQE simulations and Quantum inspired algorithm for PES analysis for many Hamiltonians. The number of Pauli coefficients these Hamiltonians have grows with the number of qubits number of qubits in their encoding. As concrete examples, we give the two smallest Hamiltonians we obtained in our experiments, at specific geometries:
\begin{itemize}
    \item $\text H_2 \text O$ defined over two qubits, at equilibrium geometry (see \Sec{sec:PESRes}):
    \begin{equation}
    \label{eq:H_H2O}
                \hat H_{\text H_2 \text O} = 0.297406 \: I \otimes Z  -0.297406 \: Z \otimes I -0.074868  \: Z \otimes Z + 0.038562 \: X \otimes X
    \end{equation}
    \item Formic Acid (see \Sec{sec:DQES} with minimal active space tested (of two qubits as well), with the geometry provided at \App{sec:formic_acid_geometry}:
\begin{equation}
        \label{eq:H_formic}
        \hat H_{HCOOH} = 0.147402 \: I \otimes Z - 0.147402 \: Z \otimes I- 0.050507  \: Z \otimes Z + 0.015736 \: X \otimes X
\end{equation}
\end{itemize} Larger Hamiltonians contain from dozens to hundreds of Pauli terms, and can be given upon request.  
\section{Electronic Structure Problem Preliminaries} \label{sec:chem_prelim}

\subsection{The Electronic Hamiltonian}
To solve the electronic structure problem for $N_n$ nuclei with positions $\mathbf{R}_1,\ldots,\mathbf{R}_{N_n}$ and $N_e$ electrons with positions $\mathbf{r}_1,\ldots,\mathbf{r}_{N_e}$, we start with the full Hamiltonian:
\begin{equation} \label{eq:full_ham}
    H_{tot} = 
    -\sum_{i=1}^{N_e} \frac{\nabla_i^2}{2} 
    -\sum_{I=1}^{N_n} \frac{\nabla_I^2}{2M_I} 
    -\sum_{i=1}^{N_e}\sum_{I=1}^{N_n} \frac{Z_I}{|\mathbf{r}_i - \mathbf{R}_I|} 
    +\sum_{i<j}^{N_e} \frac{1}{|\mathbf{r}_i - \mathbf{r}_j|} 
    +\sum_{I<J}^{N_n} \frac{Z_I Z_J}{|\mathbf{R}_I - \mathbf{R}_J|},
\end{equation}
where the terms represent electron kinetic energy, nucleus kinetic energy, electron-nucleus attraction, electron-electron repulsion, and nucleus-nucleus repulsion, respectively.

\textit{Note on Units:} Atomic units are used throughout, meaning $m_e = e = \hbar = 4\pi\epsilon_0 = 1$, where $m_e$ is the mass of the electron, $e$ is its charge, $\hbar$ is the reduced Planck constant and $4\pi\epsilon_0$ is the reciprocal of the Coulomb constant.

Under the Born-Oppenheimer approximation, the nuclei are treated as fixed point charges ($M_I \to \infty$). The electronic Hamiltonian reduces to:
\begin{equation} \label{eq:elec_ham}
    H_{elec} = - \sum_{i=1}^{N_e} \frac{\nabla_i^2}{2} - \sum_{i=1}^{N_e}\sum_{I=1}^{N_n} \frac{Z_I}{|\mathbf{r}_i - \mathbf{R}_I|} + \sum_{i<j}^{N_e} \frac{1}{|\mathbf{r}_i - \mathbf{r}_j|}
\end{equation}

\subsection{Orbitals and Fock Space}
\label{sec:fock_space}
To construct the many-body states for the $N_e$ electrons, we define a single-particle basis consisting of $N$ orthonormal spatial molecular orbitals. Associating each spatial orbital with either an up ($\uparrow$) or down ($\downarrow$) spin eigenfunction yields a set of exactly $2N$ orthonormal \textit{spin-orbitals} $\{\chi_p(\mathbf{x})\}_{p=1}^{2N}$. Here, the composite coordinate $\mathbf{x} = (\mathbf{r}, s)$ denotes both the spatial position $\mathbf{r}_i \in \mathbb{R}^3$ and the spin coordinate $s \in \{ \uparrow, \downarrow \}$.

According to the Pauli Exclusion Principle, electrons are fermions, meaning each spin-orbital can be occupied by at most one electron. In this formalism, the multi-particle states live in what is known as the \textit{Fock space}. The dimension $d$ of this subspace scales exponentially with system size and is given by $\binom{2N}{N_e}$; the number of possible configurations of $N_e$ electrons occupying $2N$ spin orbitals.
For instance, in the case where $n_e=N/2$, we have $d=\mathcal{O}(\frac{2^{2N}}{\sqrt N})$. As a concrete numerical example, for 6 electrons in 5 \textit{spatial} orbitals there are 10 spin orbitals, resulting in $d=\binom{10}{6}=210$. This corresponds to the number of what is called \textit{Fock states} or Slater determinants, which we define in the following section.
\\

We remark that the restriction to a finite number of spatial orbitals is justified by thermodynamic considerations. The occupation probability of electronic states is governed by the Fermi-Dirac distribution~\cite{ref:pathria2011statistical}. In the ground state limit ($T \to 0$), this distribution approaches a step function, where states below the chemical potential are occupied and states above it are empty. Even at finite temperatures, the probability of occupying high-energy states decays exponentially. This allows us to neglect the infinite spectrum of high-energy orbitals and treat the system using a set of $N_e/2$ spatial orbitals\footnote{That being said, basis sets often include $N>N_e/2$ spatial orbitals. The reason is that the Fermi-Dirac distribution is strictly valid only for non-interacting fermions; in reality, electron-electron repulsion (correlation) causes electrons to partially occupy higher-energy orbitals even at $T \to 0$.}.
\\
This finite discretization allows for a direct connection to quantum information. For example, a single electron confined to one spatial orbital can be treated as a theoretical realization of a qubit, where the logical states correspond to the electron's spin orientations ($\ket{\uparrow}, \ket{\downarrow}$).

\subsection{Antisymmetry and Slater Determinants}
\label{sec:SD}
Because electrons are indistinguishable fermions, the total $N_e$-electron wavefunction $\Psi(\mathbf{x}_1, \dots, \mathbf{x}_{N_e})$ must be fundamentally antisymmetric with respect to the exchange of any two particles:
\begin{equation}
\label{eq:antisym}
    \Psi(\dots, \mathbf{x}_i, \dots, \mathbf{x}_j, \dots) = - \Psi(\dots, \mathbf{x}_j, \dots, \mathbf{x}_i, \dots)
\end{equation}

A specific electronic configuration where $N_e$ electrons occupy a selected subset of the available spin-orbitals, and that explicitly satisfies the antisymmetry condition in Eq.~\ref{eq:antisym}, is described by a \textit{Slater Determinant}~$\Phi$:
\begin{equation} \label{eq:slater_det}
    \Phi(\mathbf{x}_1, \dots, \mathbf{x}_{N_e}) = \frac{1}{\sqrt{N_e!}} 
    \begin{vmatrix}
    \chi_1(\mathbf{x}_1) & \chi_2(\mathbf{x}_1) & \cdots & \chi_{N_e}(\mathbf{x}_1) \\
    \chi_1(\mathbf{x}_2) & \chi_2(\mathbf{x}_2) & \cdots & \chi_{N_e}(\mathbf{x}_2) \\
    \vdots & \vdots & \ddots & \vdots \\
    \chi_1(\mathbf{x}_{N_e}) & \chi_2(\mathbf{x}_{N_e}) & \cdots & \chi_{N_e}(\mathbf{x}_{N_e}).
    \end{vmatrix}
\end{equation}
In the formalism of second quantization, such a determinant corresponds to a unique \textit{Fock state} (or occupation number vector). 

Crucially, the set of all $\binom{2N}{N_e}$ possible Slater Determinants formed from these $2N$ spin-orbitals constitutes a complete orthonormal basis for the $N_e$-electron Hilbert space. Therefore, the exact many-body wavefunction $|\Psi\rangle$ can be expressed as a linear superposition:
\begin{equation}
    |\Psi\rangle = \sum_{k} C_k |\Phi_k\rangle,
\end{equation}
where $|\Phi_k\rangle$ represents the $k$-th Slater determinant and the coefficients satisfy $\sum |C_k|^2 = 1$. The Hartree-Fock method (discussed next) approximates this sum by finding the \textit{single} best determinant.
\\
\newline
\subsection{The Hartree-Fock Method} \label{sec:hf_method}
The Hartree-Fock (HF) theory is a method of approximation that assumes the ground state wavefunction can be described by a single Slater determinant~\cite{ref:szabo_ostlund,ref:jensen2017introduction}\footnote{Terminology varies in the literature. For instance \cRef{ref:szabo_ostlund} uses ``Hartree-Fock \textit{Approximation''}, while \cRef{ref:jensen2017introduction} refers to it as ``Hartree-Fock Theory''. We use ``Hartree-Fock \textit{Method}'' in this context.}.
It is important to emphasize that restricting the solution to a single determinant is a \textit{mean-field approximation}; it assumes each electron moves in the average static field of all others, thereby neglecting the instantaneous electron-electron correlation.

Consequently, the HF energy is necessarily higher than the exact ground state energy. Despite this limitation, the optimized HF determinant serves as the critical reference state for high-accuracy ``Post-Hartree-Fock'' methods (such as Configuration Interaction or Coupled Cluster), which aim to recover the missing correlation energy~\cite{ref:szabo_ostlund, helgaker2000molecular}.

Mathematically, we seek the Slater determinant $\Phi_{HF}$ that minimizes:
\begin{equation}
    E_{HF} = \min_{\Phi} \langle \Phi | H_{elec} | \Phi \rangle
\end{equation}
subject to the orthonormality constraints of the spin-orbitals.
This variational minimization yields the \textit{Hartree-Fock equations}:
\begin{equation}
    f(\mathbf{x}) \psi_i(\mathbf{x}) = \epsilon_i \psi_i(\mathbf{x})
\end{equation}
where $\epsilon_i$ are the orbital energies and $f(\mathbf{x})$ is the \textit{Fock operator}. This operator describes the effective Hamiltonian felt by a single electron:
\begin{equation}
    f(\mathbf{x}_i) = \underbrace{-\frac{\nabla_i^2}{2} - \sum_I \frac{Z_I}{|\mathbf{r}_i - \mathbf{R}_I|}}_{h_{core,i}} + v^{HF}(\mathbf{x}_i)
\end{equation}
Here, $h_{core}$ represents the kinetic energy and nuclear attraction, while $v^{HF}$ is the effective mean-field potential. The HF potential is defined as the sum of the Coulomb ($J$) and Exchange ($K$) operators acting over all occupied orbitals\footnote{Calculating the potential $v^{HF}$ scales as $O(N^4)$ in a single iteration due to the contraction of $O(N^4)$ two-electron integrals. This is consistent with the polynomial runtime (per iteration) discussed in Sec. \ref{sec:furtherRemarks}, distinct from the NP-complete worst-case complexity of the global optimization.}:
\begin{equation}
    v^{HF}(\mathbf{x}) = \sum_{j}^{occ} (J_j(\mathbf{x}) - K_j(\mathbf{x}))
\end{equation}
These operators are defined by their action on an arbitrary orbital $\phi(\mathbf{x})$:
\begin{itemize}
    \item \textit{Coulomb Operator ($J_j$):} Represents the classical electrostatic repulsion felt by an electron at position $\mathbf{x}$ due to the electron density in the occupied orbital $\psi_j$.
    \begin{equation}
        J_j(\mathbf{x}_1) \phi(\mathbf{x}_1) = \left[ \int d\mathbf{x_2} |\psi_j(\mathbf{x_2})|^2 \frac{1}{|\mathbf{r}_1 - \mathbf{r}_2|} \right] \phi(\mathbf{x}_1)
    \end{equation}
    
    \item \textit{Exchange Operator ($K_j$):} Represents the non-classical exchange effect arising from antisymmetry.
    \begin{equation}
        K_j(\mathbf{x}_1) \phi(\mathbf{x}_1) = \left[ \int d\mathbf{x}_2 \psi_j^*(\mathbf{x}_2) \frac{1}{|\mathbf{r}_1 - \mathbf{r}_2|} \phi(\mathbf{x}_2) \right] \psi_j(\mathbf{x}_1)
    \end{equation}
\end{itemize}
\subsection{Second Quantization and Integral Definitions}
For quantum algorithms, we project the electronic Hamiltonian (Eq.~\ref{eq:elec_ham}) onto the basis of molecular orbitals generated by the HF procedure. This yields the \textit{Second Quantized Hamiltonian}:
\begin{equation} \label{eq:second_quant}
    H = \sum_{p,q} h_{pq} a^\dagger_p a_q + \frac{1}{2} \sum_{p,q,r,s} h_{pqrs} a^\dagger_p a^\dagger_q a_r a_s
\end{equation}
where $a^\dagger_p, a_q$ are fermionic creation and annihilation operators. The coefficients are the pre-computed integrals in the molecular orbital basis:

\begin{itemize}
    \item \textbf{One-Electron Integrals ($h_{pq}$):} Corresponds to the core Hamiltonian matrix elements.
    \begin{equation}
    \label{eq:1body}
        h_{pq} = \int d\mathbf{x} \, \psi_p^*(\mathbf{x}) \left( -\frac{\nabla^2}{2} - \sum_I \frac{Z_I}{|\mathbf{r} - \mathbf{R}_I|} \right) \psi_q(\mathbf{x})
    \end{equation}
    
    \item \textbf{Two-Electron Integrals ($h_{pqrs}$):} Represents the electron-electron repulsion between electrons in MOs $p, q, r, s$. In physics notation, these are defined as:
    \begin{equation} \label{eq:2body}
        h_{pqrs} = \langle pq | rs \rangle = \int d\mathbf{x}_1 d\mathbf{x}_2 \, \frac{\psi_p^*(\mathbf{x}_1) \psi_q^*(\mathbf{x}_2) \psi_r(\mathbf{x}_1) \psi_s(\mathbf{x}_2)}{|\mathbf{r}_1 - \mathbf{r}_2|}
    \end{equation}
\end{itemize}
Note that the Coulomb ($J$) and Exchange ($K$) terms discussed in the Hartree-Fock section correspond to the specific integrals $h_{pqpq}$ and $h_{pqqp}$ respectively. The VQE Hamiltonian, however, requires the full set of general $h_{pqrs}$ integrals to describe electron correlation effects (excitations) beyond the mean-field approximation.
\section{Geometry of Formic Acid at $T=0[K]$}
\label{sec:formic_acid_geometry}
The spatial coordinates of all atoms in formic acid HCOOH in absolute zero\footnote{https://cccbdb.nist.gov/geom2x.asp}, in \AA, are:
\begin{enumerate}
    \item C - (0.0, 0.4421, 0.0)
    \item O - (-1.0461, -0.4667, 0.0)
    \item O - (1.1708, 0.1201, 0.0)
    \item H - (-0.3887, 1.4750, 0.0)
    \item H - (-0.6086, -1.3552, 0.0)
\end{enumerate}
In the future, we hope to obtain improved results for these values: First, by plotting a PES using diagonalization (as discussed in \Sec{sec:PESRes}), and later by applying the newly developed DQES method.

\section{Circuit Representation of \texttt{UCCSD} and \texttt{EfficientSU2} ansatze}
\label{sec:circuits}
For the sake of comparison, we plot the \texttt{EfficientSU2} and \texttt{UCCSD} circuits for 2 qubits.

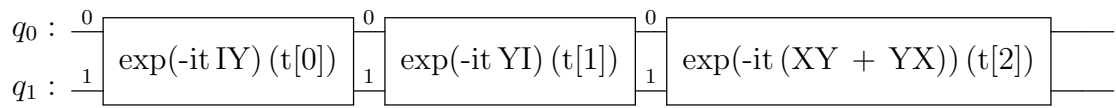
\begin{figure} [H]
    \centering
    \scalebox{1.0}{
    \Qcircuit @C=1.0em @R=1.0em @!R { \\
    	 	\nghost{{q}_{0} :  } & \lstick{{q}_{0} :  } & \multigate{1}{\mathrm{exp(\mbox{-}it\,IY)}\,(\mathrm{t[0]})}_<<<{0} & \multigate{1}{\mathrm{exp(\mbox{-}it\,YI)}\,(\mathrm{t[1]})}_<<<{0} & \multigate{1}{\mathrm{exp(\mbox{-}it\,(XY\,+\,YX))}\,(\mathrm{t[2]})}_<<<{0} & \qw & \qw\\
    	 	\nghost{{q}_{1} :  } & \lstick{{q}_{1} :  } & \ghost{\mathrm{exp(\mbox{-}it\,IY)}\,(\mathrm{t[0]})}_<<<{1} & \ghost{\mathrm{exp(\mbox{-}it\,YI)}\,(\mathrm{t[1]})}_<<<{1} & \ghost{\mathrm{exp(\mbox{-}it\,(XY\,+\,YX))}\,(\mathrm{t[2]})}_<<<{1} & \qw & \qw\\
    \\ }}
    \caption{\texttt{UCCSD} ansatz circuit for 2 qubits}
    \label{fig:UCCSD_gate}
\end{figure}
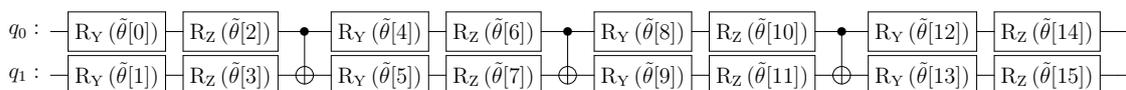
\begin{figure} [H]
    \centering
    \scalebox{0.7}{
    \Qcircuit @C=0.8em @R=0.2em @!R { 
        \nghost{{q}_{0} :  } & \lstick{{q}_{0} :  } & \gate{\mathrm{R_Y}\,(\mathrm{{\ensuremath{\vec{\theta}}}[0]})} & \gate{\mathrm{R_Z}\,(\mathrm{{\ensuremath{\vec{\theta}}}[2]})} & \ctrl{1} & \gate{\mathrm{R_Y}\,(\mathrm{{\ensuremath{\vec{\theta}}}[4]})} & \gate{\mathrm{R_Z}\,(\mathrm{{\ensuremath{\vec{\theta}}}[6]})} & \ctrl{1} & \gate{\mathrm{R_Y}\,(\mathrm{{\ensuremath{\vec{\theta}}}[8]})} & \gate{\mathrm{R_Z}\,(\mathrm{{\ensuremath{\vec{\theta}}}[10]})} & \ctrl{1} & \gate{\mathrm{R_Y}\,(\mathrm{{\ensuremath{\vec{\theta}}}[12]})} & \gate{\mathrm{R_Z}\,(\mathrm{{\ensuremath{\vec{\theta}}}[14]})} & \qw & \qw\\
        \nghost{{q}_{1} :  } & \lstick{{q}_{1} :  } & \gate{\mathrm{R_Y}\,(\mathrm{{\ensuremath{\vec{\theta}}}[1]})} & \gate{\mathrm{R_Z}\,(\mathrm{{\ensuremath{\vec{\theta}}}[3]})} & \targ & \gate{\mathrm{R_Y}\,(\mathrm{{\ensuremath{\vec{\theta}}}[5]})} & \gate{\mathrm{R_Z}\,(\mathrm{{\ensuremath{\vec{\theta}}}[7]})} & \targ & \gate{\mathrm{R_Y}\,(\mathrm{{\ensuremath{\vec{\theta}}}[9]})} & \gate{\mathrm{R_Z}\,(\mathrm{{\ensuremath{\vec{\theta}}}[11]})} & \targ & \gate{\mathrm{R_Y}\,(\mathrm{{\ensuremath{\vec{\theta}}}[13]})} & \gate{\mathrm{R_Z}\,(\mathrm{{\ensuremath{\vec{\theta}}}[15]})} & \qw & \qw\\
    }}
    \caption{\texttt{EfficientSU2} ansatz circuit for 2 qubits}
    \label{fig:SU2_gate}
\end{figure}
\section{Log-plot for 
\label{sec:logplot}
\Fig{fig:energy_comparison_10_qubit} (\texttt{UCCSD})} 
\begin{figure} [H]
    \centering
    \includegraphics[width=0.5\linewidth]{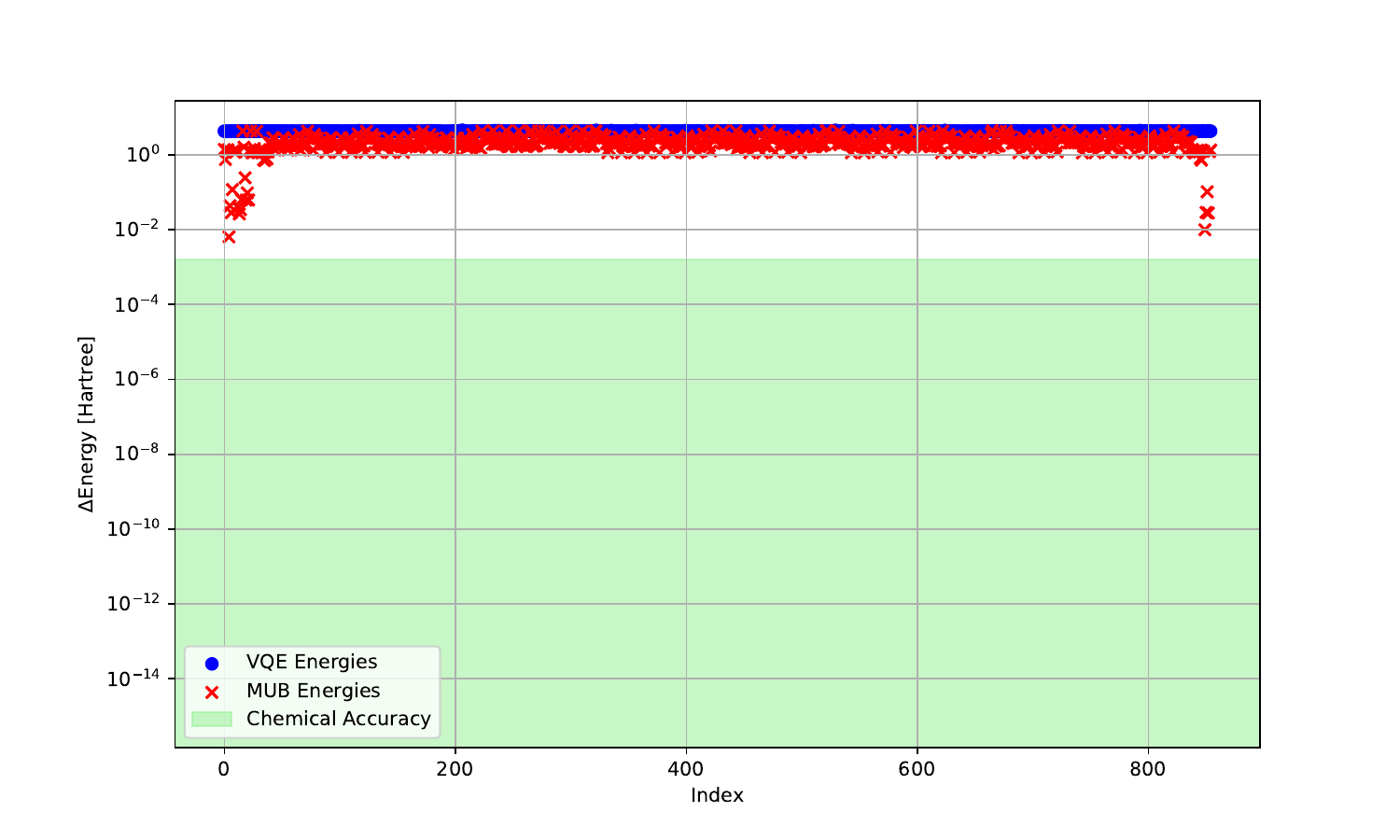}
    \caption{Log-Plot of the Energy Error spread in 10 qubits \texttt{UCCSD} in VQE trials (Blue) and MUB trials (Red).}
    \label{fig:energy_comparison_6_UCCSD_log_scale}
\end{figure}

We add the Log-plot the \texttt{UCCSD} subplot of \Fig{fig:energy_comparison_10_qubit}. This is presented in \Fig{fig:energy_comparison_6_UCCSD_log_scale}.
 The log scale in \Fig{fig:energy_comparison_6_UCCSD_log_scale} shows that the spread of energy gaps is large in DQES states, yet none of them reach chemical accuracy.
\bibliographystyle{ieeetr}
\bibliography{bib}
\end{document}